\documentclass[aps,prb,10pt,reprint,groupedaddress,superscriptaddress]{revtex4-2}

\usepackage[usenames,dvipsnames]{xcolor}
\usepackage{amssymb}
\usepackage{graphicx}
\usepackage{amsmath}
\usepackage{txfonts}
\usepackage[bookmarks=true,colorlinks,citecolor=blue,urlcolor=blue]{hyperref}
\usepackage{braket}
\usepackage{babel}
\usepackage{blindtext}
\usepackage[normalem]{ulem}

\begin{document}

\title{Kinetic Ferromagnetism and Topological Magnons of the Hole-Doped Kitaev Spin Liquid}

\newcommand{\TUM}{\affiliation{Technical University of Munich, TUM School of Natural Sciences, Physics Department, 85748 Garching, Germany}}
\newcommand{\MCQST}{\affiliation{Munich Center for Quantum Science and Technology (MCQST), Schellingstr. 4, 80799 M{\"u}nchen, Germany}}
\newcommand{\Imperial}{\affiliation{Blackett Laboratory, Imperial College London, London SW7 2AZ, United Kingdom}}

\author{Hui-Ke Jin} \TUM \MCQST
\author{Wilhelm Kadow} \TUM \MCQST
\author{Michael Knap} \TUM \MCQST
\author{Johannes Knolle} \TUM \MCQST \Imperial

\begin{abstract}
We study the effect of hole doping on the Kitaev spin liquid (KSL) and find that for ferromagnetic (FM) Kitaev exchange $K$ the system is very susceptible to the formation of a FM spin polarization. Through density matrix renormalization group simulations on finite systems, we uncover that the introduction of a single hole, corresponding to $\approx1\%$ hole doping for the system size we consider, with a hopping strength of just $t\sim{}0.28K$ is enough to disrupt fractionalization and polarize the spins in the [001] direction due to an order-by-disorder mechanism. Taking into account a material relevant FM anisotropic spin exchange $\varGamma$ drives the polarization towards the [111] direction via a reorientation  transition into a topological FM state with chiral magnon excitations. We develop a parton mean-field theory incorporating fermionic holons and bosonic spinons/magnons, which accounts for the doping induced FM phases and topological magnon excitations. We discuss experimental signatures and implications for Kitaev candidate materials. \end{abstract}

\maketitle

\noindent \textbf{INTRODUCTION}

\noindent Kinetic ferromagnetism, resulting from the subtle interplay between the motion of electrons and their interactions, provides a counter-intuitive example of a strong interaction effect in condensed matter physics~\cite{Auerbach1998,Mattis2006}. Nagaoka~\cite{Nagaoka1966} famously showed how the interference of paths from a single hole doped into a half-filled Hubbard model with infinite repulsion $U$ can lead to a ferromagnetic (FM) state in a system that typically supports antiferromagnetic (AFM) order. 
Given that the required large interaction limit is an experimental challenge,  signatures of Nagaosa's ferromagnetism have only recently been observed experimentally in quantum dots~\cite{dehollain2020nagaoka} and semiconductor heterostructures~\cite{Ciorciaro2023}.
In this context, intriguing questions are whether similar kinetic magnetism can appear in other correlated models that are experimentally accessible; and whether the kinetic FM state itself can be non-trivial, e.g., host chiral excitations.

Seemingly unrelated exotic phases of matter are quantum spin liquids (QSLs) characterized by long-range quantum entanglement and fractionalized spin excitations~\cite{Anderson73,Anderson87,Lee08,Balents10,Savary2016,QSLRMP,Knolle2019,Broholm2020}. The Kitaev honeycomb model is a celebrated example of a two-dimensional $Z_2$ QSL~\cite{Kitaev06}. The famous exact solution shows that the Kitaev spin liquid (KSL)~\cite{Kitaev06} has a gapless ground state with Dirac-type Majorana excitations, while its static $Z_2$ flux excitations remain gapped. 
In recent years, the Kitaev honeycomb model has gained experimental relevance with several spin-orbit-coupled $4d$ and $5d$ transition metal candidate materials, such as Na$_2$IrO$_3$~\cite{Choi2012,Cao2013,Manni2014,Takagi2019} and $\alpha$-RuCl$_3$~\cite{Plumb2014,Sears2015,Sandilands2015,Banerjee2016,Do2017,Baek2017,Zheng2017}, having been proposed as promising realizations of Kitaev's bond-anisotropic Ising interaction ($K$)~\cite{Jackeli2009,Chaloupka2010,Katukuri2014,Rau2014,Kee2016,Trebst2017,Winter2016,Winter2017,Winter2017NC,Knolle2018}. However, the KSL is fragile and residual zigzag magnetic order appears in all materials at low temperatures~\cite{Chaloupka2013,Yamaji2016}, because of the inevitable presence of additional interactions such as the off-diagonal symmetric ($\varGamma$) and Heisenberg ($J$) exchanges~\cite{Plumb2014,Rau2014}. Hence, understanding the robustness of the KSL in the extended $K$-$\varGamma$-$J$ Kitaev model has become a subject of intense research~\cite{Rau2014,Song2016,Gohlke2017,Gohlke2020,Wang2019,Buessen2021,Zhang2021,Li2022,Chaloupka2015}, e.g. the $K$-$\varGamma$ (with $J=0$) is believed to remain disordered~\cite{Rau2014, Janssen2017, Catuneanu2018, Gohlke2018}. 
Here, we explore a different route by studying the hole doping resilience of the KSL and uncover a surprising connection between the KSL and kinetic ferromagnetism. 

\begin{figure}[tb]
	\includegraphics[width=1.0\linewidth]{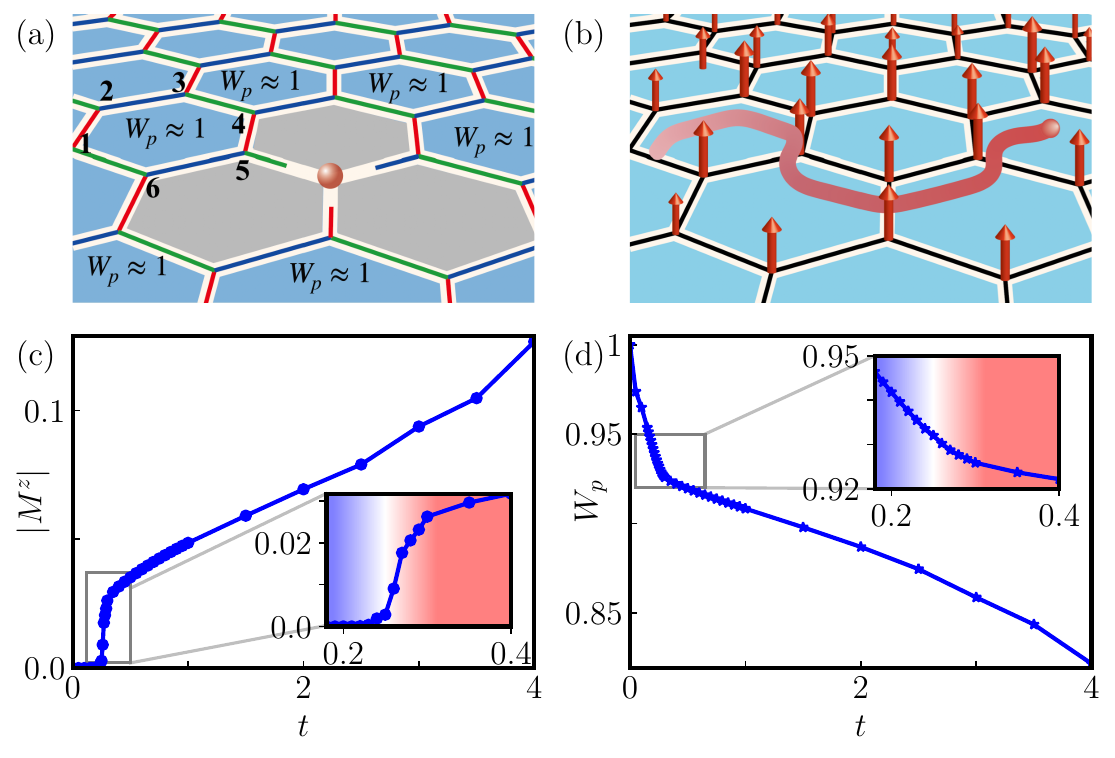}
	\caption{Magnetization in the hole-doped Kitaev spin liquid (KSL). (a) Illustration of the $t$-$K$ model with three types of bonds, $\gamma=x,y,z$. Sites $j=1,...,6$ label the convention for plaquette operators $W_p$. The system sustains the site-diluted KSL ground state with fractionalized spins ($W_p\approx{}1$) only for a slow hole (red ball). (b) For $t\gtrapprox{}K$ doping depletes plaquette operators $W_p$ while simultaneously polarizing the spins, which leads to FM order along the [001] direction. (c,d) The ground-state phase diagram for the one-hole doped $t$-$K$ model ($K=1$) on a YC4 cylinder with length $L_x=12$ (corresponding to a doping level of $\delta\approx{}0.01$). The evolutions of (c) the magnetic moment $M^z$ and (d) the average flux $W_p$ indicate a phase transition from a site-diluted KSL (blue regime in insets) to a FM state (red regime) around $t^*\approx{}0.28$. }\label{fig:fig1}
\end{figure}

In this work, we account for the effects of hole doping by considering an extension of the Kitaev honeycomb model, referred to as the $t$-$K$ model, where $t$ denotes the hopping strength of doped holes. In the slow hole limit ($t\ll{}K$), the KSL phase is robust and holes only introduce quasi-static vacancies~\cite{Willans2011,Chalker2014}, see Fig.~\ref{fig:fig1}(a). However, in realistic systems, $t$ is typically much larger than the Kitaev exchange $K$, and previous works based on parton mean-field theories suggested superconducting ground states in the hole-doped regime~\cite{You2012,Hyart2012,Okamoto2013,Scherer2014}. Yet, our numerics does not support superconductivity at the considered low doping values, see Supplementary Note 2.  Recently, a density matrix renormalization group (DMRG) study suggested a charge density wave ground state emerging on top of a hole-doped AFM KSL, in which the superconducting correlations fall off almost exponentially at long distances~\cite{Peng2021}.

We investigate the ground state of the $t$-$K$ model by DMRG~\cite{White1992,White1993} and show that the FM KSL is remarkably fragile already for small and slow hole doping. For hopping strengths $t$ of the order of the KSL's flux gap,  $\Delta_{}E_{\rm v}\sim0.1K$, the system is already partially FM polarized by the itinerant holes, spontaneously breaking the time-reversal symmetry, as illustrated in Fig.~\ref{fig:fig1}(b). To account for our numerical results, we develop a parton theory incorporating fermionic holes and bosonic spinons/magnons. It unveils that the hole kinetic term effectively serves as a FM Heisenberg coupling destabilizing the KSL. In addition, our parton theory shows that the resulting FM order along the [001] direction originates from an order-by-disorder mechanism~\cite{Shender1982,Henley1989}. We further uncover that the presence of a finite FM off-diagonal exchange, $\varGamma>0$, shifts the magnetization direction from [001] to [111]. Remarkably, due to the change of spin polarization direction, the kinetic FM state becomes non-trivial. It spontaneously forms topological FM order with chiral magnon edge modes, akin to the FM state of the Kitaev model induced by a strong external magnetic field~\cite{Joshi2018,McClarty2018}. Furthermore, we show that hole doping can significantly lower the critical field above which the field-polarized FM state appears in Kitaev candidate materials, such as $\alpha$-RuCl$_{3}$. \\


\noindent \textbf{RESULTS}

\noindent \textbf{Model Hamiltonian}

\noindent 
The celebrated Kitaev honeycomb model is defined by the following Hamiltonian
\begin{equation}
	\mathcal{H}_{\rm S} =\sum_{\langle{}jk\rangle\in{}\gamma{}}H^{\gamma}_{jk},~~~
	H_{jk}^\gamma=-KS^\gamma_jS^\gamma_k,\label{eq:HS}
\end{equation}
where $S^\gamma_j$ ($\gamma=x,y,z$) are the $S=1/2$ spin operators and $S^\gamma_jS^\gamma_k$ are Ising couplings according to the $\gamma$-type of nearest-neighbor (NN) $\langle{}jk\rangle$ bonds.
Microscopic derivations~\cite{Winter2016,Winter2017,Banerjee2017,Wang2017,Ran2017} have shown that the Kitaev interaction of several material candidates is likely FM. Therefore, we focus on positive $K$ and set $K=1$ as the unit of energy. 
There exist conserved plaquette operators $W_p$ on each hexagon $p$ as $$W_p\equiv{}2^6S^x_1S^y_2S^z_3S^x_4S^y_5S^z_6,$$ where the lattice sites $j=1,...,6$ correspond to the elementary hexagon plaquette $p$ shown in Fig.~\ref{fig:fig1}(a). 
The pristine KSL ground state has $W_p=1$ for all plaquettes. 

To describe the physics of hole doping, we introduce the $t$-$K$ model~\cite{Shitade2009,You2012,Laubach2017} 
\begin{eqnarray}
	\mathcal{H}=-t\sum_{s,\langle{}jk\rangle}\left(\mathcal{P}c^\dagger_{j,s}c^{}_{k,s}\mathcal{P}+h.c.\right)+\mathcal{H}_{\rm S}, \label{eq:H}
\end{eqnarray}	
where $c^\dagger_{s}$ is the creation operator of an electron with spin index $s=\uparrow,\downarrow$ and the projector $\mathcal{P}$ removes doubly occupied states. Spin operators, $S^\gamma_j={\bf c}^\dag_j\sigma^\gamma{\bf c}^{}_j/2$, are given by the fermionic vectors ${\bf c}^\dag_j=(c^\dag_{j,\uparrow},c^\dag_{j,\downarrow} )$ and the standard $\sigma^\gamma$ Pauli matrices. The hole doping is parameterized by $\delta$ such that $\sum_{s}\langle{}c^\dag_{j,s}c^{}_{j,s}\rangle=1-\delta$. Note that the NN hopping is spin-independent and the effects of spin-orbital coupling are nevertheless retained in $\mathcal{H}_{\rm S}$~\cite{Shitade2009}, resulting in a space symmetry group of $D_{3d}$. Besides, the $t$-$K$ model is also symmetric under time reversal transformation and preserves the charge-U(1) conservation.

In order to implement DMRG calculations, the system is placed on a two-dimensional cylindrical geometry, dubbed YC$L_y$, with periodic boundary conditions (PBCs) along the short direction (with $L_y$ unit cells), while the longer one ($L_x$) is open. See details about the honeycomb lattice in Supplemental Note 1.\\

\noindent {\bf Phase Diagram}

\noindent First, we focus on the one-hole doped system.  The ground-state phase diagram on YC4 cylinders is analyzed by studying the average magnetic moments $\bm{M}=\langle{}\bm{S}_i\rangle$ and the average $Z_2$ flux $W_p$, as shown in Fig.~\ref{fig:fig1}(c,d). For $t<t^*$, the ground state corresponds to a site-diluted KSL, where the kinetic energy of the holes is insufficient to overcome the $Z_2$ flux excitation gap $\Delta{}E_{v}\sim{}0.1K$. Instead, the presence of the slow-moving holes can be thought of as quasi-static vacancies within the KSL state~\cite{Willans2011,Chalker2014}. Although $W_p$ slightly decreases as $t$ increases in this phase, the system does not acquire any magnetization.  

We observe that a phase transition occurs at $t=t^*\approx{}0.28K$, as visible by kinks in the curves of both the magnetization and $Z_2$ flux, see inset plots of Figs.~\ref{fig:fig1}(c) and (d). As one of our main findings, already for $t>t^*$ a FM phase appears with a finite magnetization along the [001] direction. We note that the first-order derivative of the ground-state energy also exhibits a kink around $t^*\approx{}0.28{}K$ which suggests that the phase transition is of second order, see Supplementary Note 2.

This FM phase is an example of kinetic magnetism, which spontaneously breaks the time-reversal symmetry and arises solely from the holes' kinetic energy. Remarkably, the KSL correlations are only partially depleted, as signaled by the gradual decrease of the $Z_2$ flux $W_p$ in Fig.~\ref{fig:fig1}(d). This coexistence of magnetization and finite plaquette flux is a general feature of the [001] FM phase until the system is fully polarized by a sufficiently large hopping strength. 
According to Nagaoka's theorem, for a single-hole doped system~\cite{Nagaoka1966} in the thermodynamic limit, the saturated FM state is only reached for vanishing spin exchange (or in other words, the on-site Hubbard $U\rightarrow\infty$). Therefore, the intriguing coexistence of FM order and KSL correlations could generally persist in systems described by the $t$-$K$ model.

\begin{figure}[tb]
	\centering
	\includegraphics[width=1.0\linewidth]{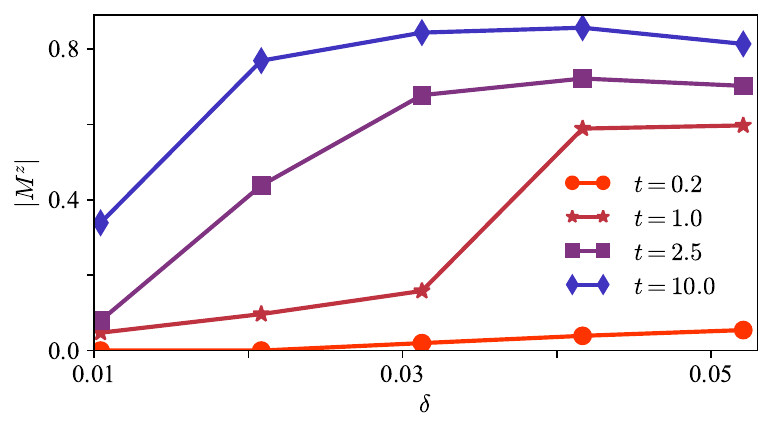}
	\caption{The magnetization as a function of doping level $\delta$ for variable values of hopping strength on YC4 cylinders with length $L_x=12$ and $K=1$.}
	\label{fig:Mz_delta}
\end{figure}

We further investigate whether FM order persists for multiple holes and study the magnetization as a function of hole density $\delta$ in Fig.~\ref{fig:Mz_delta}. Unlike Nagaoka's ferromagnetism, which may disappear for a thermodynamic density of holes~\cite{Putikka1992}, we find that the FM order observed in our work is robust and can be further enhanced with multi-hole doping. First, a larger $\delta$ can lower the critical value of $t^*$. For instance, at $t=0.2K$, the system remains in the site-diluted KSL phase without any magnetization for $\delta\lesssim0.02$, while a small but finite magnetic order emerges for $\delta>0.02$. For the hopping strength $t$ which is already capable of polarizing spins at $\delta\approx0.01$, a slight increase in $\delta$ can significantly enhance the magnetization until it reaches a saturation value at $\delta\approx0.05$. Due to a rapid increase in entanglement entropy, our DMRG simulations face convergence problems for systems with even higher doping levels. Nevertheless, our results suggest that the ferromagnetic order is a generic feature of the $t$-$K$ model, at least in the low doping limit $\delta\leq{}0.06$.\\

\noindent {\bf Off-Diagonal Exchanges}

\noindent 
Next, we study the hole-doped extended Kitaev honeycomb model, dubbed the $t$-$K$-$\varGamma$ model, as a more realistic system that incorporates additional off-diagonal symmetric spin exchange parameterized by $\varGamma$. The spin part of the model now becomes
\begin{equation}
	\mathcal{H}_{\rm S} =\sum_{\langle{}jk\rangle\in{}\gamma{}}H^{\gamma}_{jk},~~~
	H_{jk}^\gamma=-KS^\gamma_jS^\gamma_k-\varGamma{}\left(S^\alpha_jS^\beta_k+S^\beta_jS^\alpha_k\right), \label{eq:HS2}
\end{equation}
with $\alpha\neq{}\beta\neq{}\gamma$ for the $\varGamma$ terms on $\gamma$-type bonds. Although the $\varGamma$ term is typically believed to be AFM~\cite{Janssen2017,Cookmeyer2018}, here we focus on the effect of a FM $\varGamma>0$ as this has the most interesting consequences. 

\begin{figure}[tb]
	\includegraphics[width=1.0\linewidth]{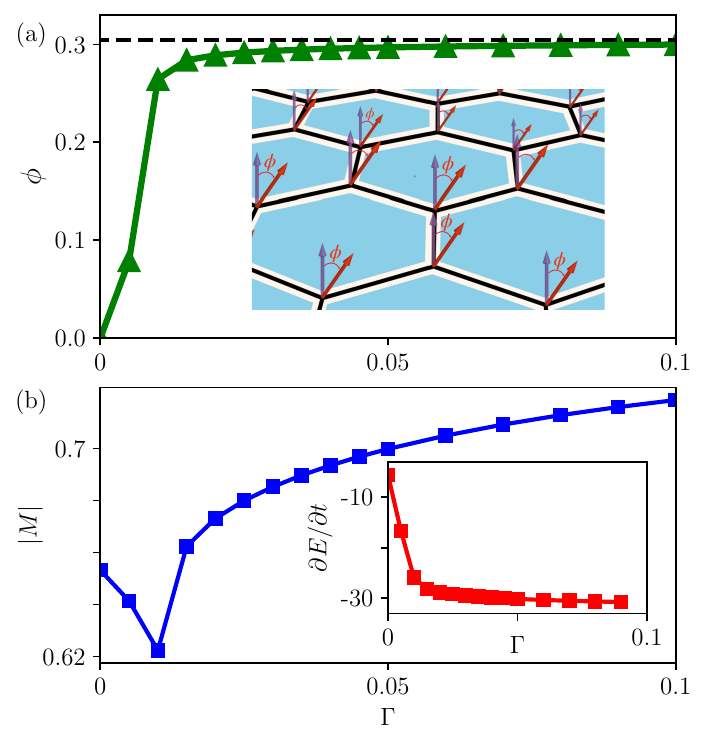}
	\caption{Effect of ferromagnetic off-diagonal symmetric exchange $\varGamma$. (a) The polar angle $\phi\equiv{}\cos^{-1}\left(M^z/|M|\right)$ representing the magnetization direction as a function of $\varGamma$, where $\phi=0$ indicates the [001] direction and the black dashed line the [111] direction. Inset: a finite FM $\varGamma>0$ can induce a rotation of the spin polarization direction from [001] (transparent purple arrows) to [111] (solid red arrows) by an angle $\phi$.  (b) The amplitude of magnetization as a  function of $\varGamma$ for the one-hole doped model with $t=20K$ on a YC4 cylinder with length $L_x=12$, corresponding to doping level $\delta\approx0.01$. Inset: The first-order derivatives of ground-state energy versus $\varGamma$. }\label{fig:effectGamma}
\end{figure}

In the absence of $\varGamma$, the spin polarization is always along the [001] direction (or by symmetry equivalently, the [010] and [100] directions), which can be attributed to an order-by-disorder mechanism as we will explain below. In Fig.~\ref{fig:effectGamma}(a), we show that the presence of $\varGamma>0$ can shift the spin polarization direction from [001] to [111]. Indeed, already a tiny value of $\varGamma^*$ is sufficient to induce a significant change in the polar angle of magnetization, denoted as $\phi\equiv{}\cos^{-1}(M^z/|M|)$. For $\varGamma=0.02$, the polar angle is already very close to the value of $\phi_{[111]}=\cos^{-1}(1/\sqrt{3})$, which corresponds to a FM order along the [111] direction, see Fig.~\ref{fig:effectGamma}(b). 
At $\varGamma^*\approx{}0.01K$, a noticeable kink in the first-order derivative of the ground-state energy in Fig.~\ref{fig:effectGamma}(c) indicates that this change in spin polarization direction is accompanied by a phase transition.
We will discuss below that this transition leads to topologically non-trivial magnon excitations. 
The magnetization displays a dip around $\varGamma^*$, corroborating the findings from the analysis of the ground-state energy. Note that a moderately large $\varGamma$ can also enhance the magnitude of the magnetization.

Compared to the FM $\varGamma$, the perturbation of an AFM $\varGamma$ has a much smaller impact on the [001] ordered ground state. For instance, a small AFM $\varGamma\approx{}-0.02$ only induces a canting field in the [001] order. However, a FM $\varGamma$ with a similar magnitude is sufficient to shift the polarization direction from [001] to [111]. \\

\noindent {\bf Parton mean-field theory}

\noindent 
In order to understand the selection of the FM moment direction and the impact of $\varGamma$, we develop a parton mean-field theory. To effectively describe the low-energy degrees of freedom in the kinetic FM phase, we consider the following parton representation for electron operators~\cite{Jayaprakash1989} with $c^\dag_{j,s}\equiv{}h^{}_{j}b^\dag_{j,s}$. Here, $b^\dag_{j,s}$'s are Schwinger boson operators representing spinons, and $h^{}_{j}$'s are fermionic holon operators representing empty sites. The local constraint, $h^\dag_jh^{}_j+\sum_{s}b^\dag_{j,s}b_{j,s}=1$,  ensures the original three-dimensional physical local Hilbert space. The parton theory is constructed such that it allows for spontaneous symmetry breaking toward ordered states.

The $t$-$K$-$\varGamma$ model takes the form
\begin{equation}
	\mathcal{H}=-t\sum_{\langle{}jk\rangle}\left(\hat{F}_{jk}h_k^\dagger{}h^{}_j+h.c.\right)+\sum_{\langle{}jk\rangle\in\gamma}h^{}_{j}h^\dag_{j}h^{}_{k}h^\dag_{k}\hat{H}_{jk}^\gamma{},\label{eq:Hparton}
\end{equation}
with $\hat{F}_{jk}\equiv{}\sum_{s}b^\dag_{j,s}b^{}_{k,s}$ representing short-range FM spin correlations~\cite{Arovas1988}. One can observe that the hopping term in Eq.~\eqref{eq:Hparton} connects the hole's kinetic energy $\hat{P}_{jk}\equiv{}h^\dag_{j}h^{}_k$ with FM correlations $\hat{F}_{jk}$, suggesting that finite doping with dominant $t$ dramatically renormalizes the spin-spin interactions and fosters FM order.

The second term in Eq.~\eqref{eq:Hparton}, $\hat{H}_{jk}^\gamma$, represents spin interactions as defined in Eq.~\eqref{eq:HS2}, but expressed in terms of bosonic spinons instead of electron operators. We develop a large-$N$ mean field theory within a Schwinger-boson approach~\cite{Arovas1988,Auerbach1988,Sachdev1992} in which  the spinon occupancy is given by $n_b=\sum_s b^\dag_{j,s}b^{}_{j,s}=1-\delta$ (see Supplementary Note 3).

Guided by our DMRG results, our interest lies in the FM phase with $t\gg{}K$. We also explored the possibility of a gapped QSL phase but found no evidence thereof in the limit of $n_b\rightarrow~1$. Hence, the most natural ansatz is a FM state along the general direction of  $(\sin\phi\cos\theta, \sin\phi\sin\theta, \cos\phi)$. For conveniently describing magnon excitations, we can replace the Schwinger bosons with Holstein-Primakoff (HP) bosons as~\cite{Holstein1940,Auerbach1998}
\begin{equation*}
	(b^{}_{j,\uparrow},b^{}_{j,\downarrow})^T\rightarrow{}e^{-i\theta\sigma^z/2}e^{-i\phi\sigma^y/2}(\sqrt{n_b-a^\dag_ja^{}_j}, a^{}_j)^T,
\end{equation*} 
where $a^{\dag}_j$ and $a^{}_j$ are the creation and annihilation operators for the HP bosons (magnons), respectively. 
Then we perform a standard Hartree-Fock decoupling of Eq.~\eqref{eq:Hparton} to obtain a mean-field theory with separate hole and spin parts, dubbed $\mathcal{H}_{\rm h}$ and $\mathcal{H}_{\rm s}$, respectively. Here, $\mathcal{H}_h$ refers to a spinless free-fermion band with renormalized bandwidth by the factor $|\langle\hat{F}_{jk}\rangle|$ and filling of $\delta$. The spin part $\mathcal{H}_{\rm s}$ is treated in spin-wave theory (SWT) incorporating up to fourth-order Holstein-Primakoff expansions, and the spin amplitude is renormalized to $S=n_b/2$. The expectation values of FM spin correlations $\hat{F}_{jk}$ and hole's kinetic energy $\hat{P}_{jk}$ are determined self-consistently. Remarkably, we can now see that for $\delta\ll{}1$ the kinetic energy effectively acts as a {\rm FM} Heisenberg interaction with a coupling constant $J\equiv{}-\langle{}P_{jk}\rangle{}t/n_b\sim{}|t|\delta$. 

The semi-classical ground-state energy of the FM Kitaev honeycomb model ($t=\varGamma=0$) turns out to be independent of $\phi$ and $\theta$, yielding an emergent $O(3)$ manifold of degenerate ground states. This classical $O(3)$ degeneracy can be lifted by quantum fluctuations in a quantum order-by-disorder mechanism~\cite{Shender1982,Henley1989}. Within SWT the quantum correction to the ground-state energy is  $E_{s}=-\kappa(\kappa+2)\mathrm{N}_cK/4+3\kappa(2\kappa+1)\mathrm{N}_cJ+\sum_{{\bf k},n}\omega_{{\bf k},n}$, where $\mathrm{N}_c$ is the number of the unit cells and $\omega_{{\bf k},n}$ are the two magnon bands of the honeycomb lattice. Note that $\omega_{{\bf k},n}$ depend on $\phi$ and $\theta$ implicitly, and thereby so does $E_{s}$. Our calculations reveal that a magnetization along the [001] direction (or equivalently, the [100] and [010] directions) is energetically preferred in accordance with our DMRG results (see Supplementary Note 3).

A complication arises from the fact that the semi-classical ground state of the pure Kitaev honeycomb model is a classical spin liquid~\cite{Samarakoon2017} which is manifest in the {\it linear} SWT for arbitrary spin polarization directions as a nearly flat magnon band with almost zero energy (see Supplementary Note 3). Thus, an HP expansion that only includes the quadratic terms is insufficient to correctly capture the quantum fluctuations. We employ a {\it nonlinear} SWT by including the quartic terms in the HP expansion and treat these within the Hartree-Fock decomposition~\cite{Chernyshev2009}. We indeed find significant renormalizations of the flat magnon bands, for instance, a closing of the gap between formerly flat magnon bands in the nonlinear SWT for $\varGamma=0$. More technical details can be found in Supplementary Note 3.


Including a FM $\varGamma>0$ breaks the semi-classical $O(3)$ degeneracy of the ground state. Our SWT shows that the semi-classical ground state has [111] magnetic order rather than [001], which is again consistent with our DMRG calculations, see Fig.~\ref{fig:effectGamma}(b). From the magnetic field polarized regime of the Kitaev model, it is known that different polarization directions can have qualitatively different types of boundary magnon excitations as first pointed out in Ref.~\cite{Joshi2018,McClarty2018}. For spins polarized along the [001] direction with $\varGamma=0$, the two magnon bands touch linearly and are almost non-dispersive along one of two primitive vectors. The system thereby cannot support nontrivial boundary excitations. On the other hand, the introduction of a finite $\varGamma>0$ modifies the magnon bands from the change of the spin polarization to the [111] direction. Consequently, the two magnon bands are separated by a gap and acquire non-zero Chern numbers $(-1, 1)$ for the lower and upper magnon bands, respectively~\cite{TKNN,Fukui2005,Joshi2018}. Hence, as our second main result, we find that the kinetic FM state of the doped Kitaev model (with small $\varGamma>0$) spontaneously breaks TRS forming a topological magnon insulator. 
This explains the phase transition found when turning on the off-diagonal $\varGamma$ interactions, see Fig.~\ref{fig:effectGamma}.

\begin{figure}[tb]
	\centering
	\includegraphics[width=1.0\linewidth]{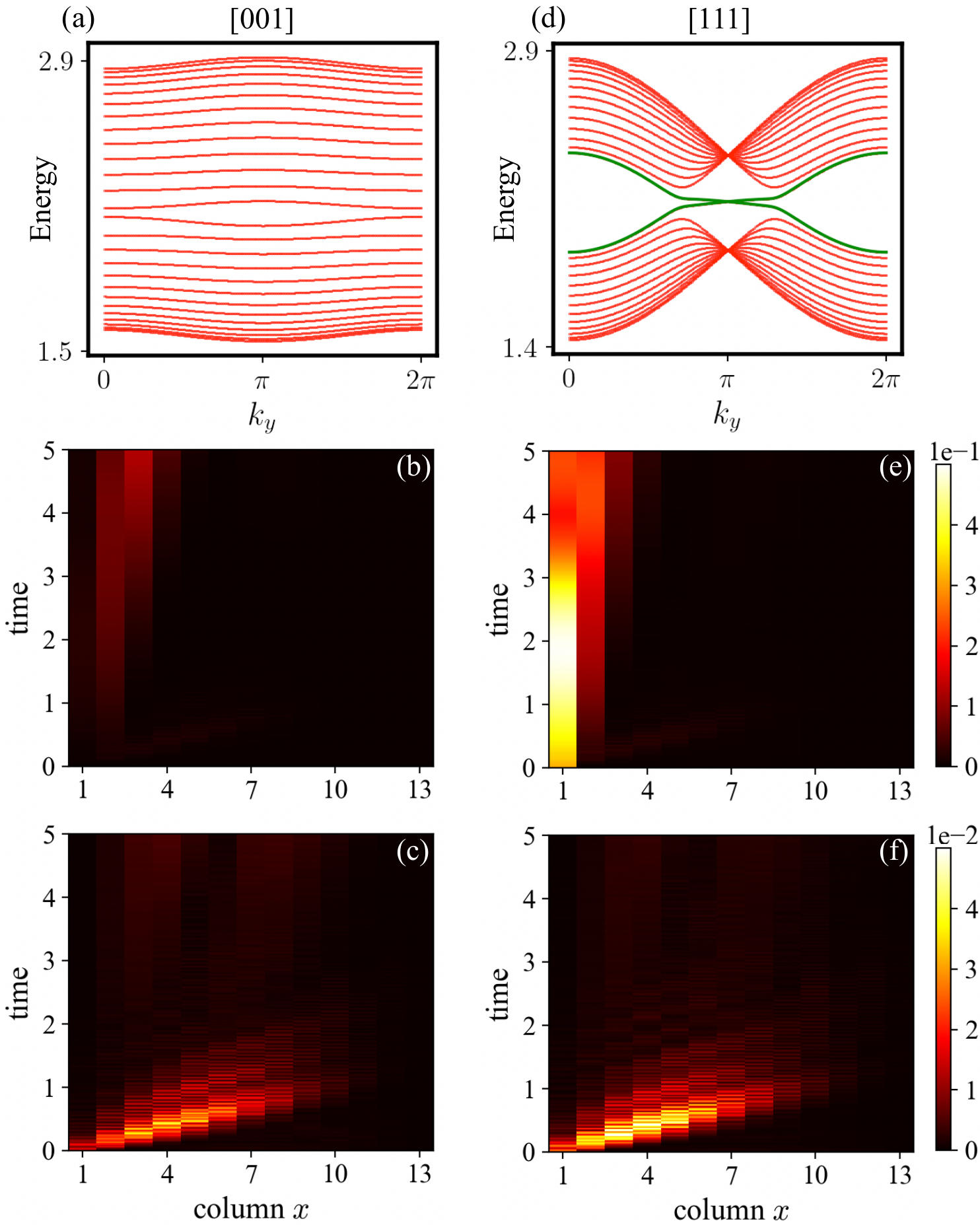}
	\caption{Distinct boundary magnon excitations for the [001] and [111] polarized FM orders. (a) Magnon bands on a cylinder for a [001] ordered state within nonlinear SWT. (b, c) The real-time dynamics of (b) spin current $j_{s}(t,x)$ and (c) charge current $j_{c}(x,t)$ for a [001] ordered state. These results are obtained by DMRG using one-hole doped YC3 cylinders ($\delta\approx0.013$). For (a, b, c), we set the off-diagonal exchange $\varGamma=0$ and hopping strength $t=10K$. The self-consistent solution with doping level $\delta=0.01$ gives kinetic hole energy $\langle{}\hat{P}_{jk}\rangle\approx-0.93\times{}10^{-2}$ and FM spin correlation $\langle{}\hat{F}_{jk}\rangle\approx0.79$ on the $x-$ ($y$-) type bonds as well as $\langle{}\hat{F}_{jk}\rangle\approx0.82$ on the $z$-type bonds. 
		(d) Magnon bands within nonlinear SWT on a cylinder for the [111] ordered state. The chiral edge states are marked in green. (e, f) The real-time dynamics of (e) spin current $j_{s}(t,x)$ and (f) charge current $j_{c}(x,t)$ for a [111] ordered states within DMRG. For (d, e, f), we set $\varGamma=0.05K$ and $t=10K$. The self-consistent solution with $\delta=0.01$ gives kinetic hole energy $\langle{}\hat{P}_{jk}\rangle\approx-0.93\times{}10^{-2}$ and FM spin correlation $\langle{}\hat{F}_{jk}\rangle\approx0.85$ on all three type bonds. }\label{fig:realtime}
\end{figure}

To see how the change in spin polarization can give rise to chiral boundary states, we calculate the magnon bands within SWT on narrow cylinders. With PBCs along $y$-direction, $k_y$ is still a good quantum number and the 2D systems can be treated as $L_y$ independent 1D chains. The magnon spectra for the [001] and [111] ordered states are shown in Figs.~\ref{fig:realtime}(a) and (d). Indeed, only the [111] ordered state has chiral boundary modes (green) as predicted. \\

\noindent  {\bf Real-time dynamics}

\noindent 
So far our observation that a topological FM state is realized for the [111] direction solely relies on the parton+SWT description. A numerical confirmation by DMRG is complicated by the fact that the ground state is topologically trivial but only the magnon excitation spectrum carries signatures of the magnon Chern bands. Therefore, we study the distinct real-time dynamics of excitations in the [001] and [111] ordered FM states. To obtain an initial state with a magnon-like excitation at the boundary, we apply a local time-reversal operator at lattice site $j$, which couples to spin excitations for arbitrary spin polarizations, to a one-hole doped ground state $|\Psi\rangle$ as  
\begin{equation}
	|\Psi(j)\rangle\equiv{}i\sigma^y_{j}\mathcal{K}|\Psi\rangle, \label{eq:Psij}    
\end{equation} 
where $\mathcal{K}$ denotes the complex conjugate operator. Note that $\sigma^y_{j}$ also annihilates the components of $|\Psi\rangle$ which contain a local hole at site $j$. As a result, in addition to flipping the spins at site $j$, it also weakly excites charge excitations and slightly modifies the global magnetization pattern. Starting with the prepared initial state $|\Psi(j)\rangle$, the system is evolved by performing a standard real-time evolution as $|\Psi(j,t)\rangle=e^{-i\mathcal{H}t}|\Psi(j)\rangle$. In practice, we approximately represent the short-time unitary operator, $e^{-i\mathcal{H}\mathrm{d}t}$, as a compact matrix product operator (MPO)~\cite{Zaletel2015}, where a small time step ${\rm d}t=0.01K$ has been used. Then the time evolution can be efficiently simulated by applying such an MPO to $|\Psi(j,t)\rangle$ successively.

We create a one magnon defect at site $j$ in the leftmost column of the YC cylinders and follow the spin current for each column $x$ defined as
\begin{equation}
	j_{s}(x,t)=\sum_{y}|\mathrm{d}\langle\bm{S}_{x,y}(t)\rangle/\mathrm{d}t|, \label{eq:js} 
\end{equation}
with $\langle{}..\rangle$ the expectation value with respect to $|\Psi(j,t)\rangle$ and $(x, y)$ denoting the lattice coordinates. Note that $j_{s}(x,t)$ is obtained by summing over all the sites belonging to the $x$-th column, namely, summing over allowed $k_y$. Additionally, we also keep track of the charge current
\begin{equation}
	j_{c}(x,t)=\sum_{y}\mathrm{d}\langle{}n^{h}_{x,y}(t)\rangle/\mathrm{d}t, \label{eq:jc}        
\end{equation}
where $n^h_{j}=1-\mathcal{P}\sum_{s}c^{\dag}_{j,s}c^{}_{j,s}\mathcal{P}$ is the occupancy of holes. 

The simulations for the dynamics are performed on YC3 cylinders, and the results are shown in Fig.~\ref{fig:realtime}. We observe a distinct difference in the spin currents between the [001] and [111] ordered states, as evidenced by the leftmost boundary columns in Figs.~\ref{fig:realtime}(b) and \ref{fig:realtime}(e), respectively. The spin current in the [001] ordered state shows very little dynamics, as indicated by its small magnitude of $j_{s} \sim 10^{-4}$ in Fig.~\ref{fig:realtime}(b).  It can be attributed to the absence of distinct boundary modes and the almost vanishing velocity of magnon modes. In stark contrast, for the [111] direction $j_{s}(x,t)$  exhibits a strong and long-live signal in the leftmost boundary as expected for a localized surface excitation. Note that Eq.~\eqref{eq:Psij} unavoidably involves excitations related to bulk magnon bands, which explains why the boundary spin current can leak into the bulk. We have checked that the magnetic field-induced topological FM in the [111] direction~\cite{Joshi2018,McClarty2018} shows a very similar response confirming the presence of chiral magnons. 

The charge current for the [001] and [111] ordered states are very similar, see Figs.~\ref{fig:realtime}(c) and (f), respectively. This is consistent with our parton mean-field theory. Both the [001] and [111] ordered states give rise to similar short-range FM spin correlations (see Supplementary Note 3),  which suggests that the holon Hamiltonians are similar.\\

\noindent {\bf Implications for the field-polarized state of $\alpha$-RuCl$_3$}

\noindent  
Unlike the simplified models introduced in Eq.~\eqref{eq:HS} and Eq.~\eqref{eq:HS2}, more realistic models for experimentally relevant materials are much more complicated. Here, we focus on $\alpha$-RuCl$_3$ and related materials~\cite{Winter2017,Trebst2017,maksimov2020rethinking} which are often described by a number of parameters, including $\{K,J,\varGamma,\varGamma',J_{3}\}$. The corresponding Hamiltonian is expressed as 
\begin{equation}
\begin{split}
\mathcal{H}_{\rm S}^{\alpha-RC_3}=&\sum_{\langle{}ij\rangle{}\in\gamma}\left[-KS^\gamma_jS^\gamma_{k}-\varGamma\left(S^\alpha_jS^\beta_{k}+S^\alpha_jS^\beta_{k}\right)\right]+\\
&\sum_{\langle{}ij\rangle{}\in\gamma}-\varGamma'\left(S^{\gamma}_iS^{\alpha}_j+S^{\gamma}_iS^{\beta}_j+S^{\alpha}_iS^{\gamma}_j+S^{\beta}_iS^{\gamma}_j\right)+\\
&-\sum_{\langle{}ij\rangle{}}J\vec{S}_i\cdot\vec{S}_j-\sum_{\langle{}ij\rangle_3}J_3\vec{S}_i\cdot\vec{S}_j,
\end{split}\label{eq:HaRC3}
\end{equation}
where $\alpha\neq\beta\neq\gamma$ and $\langle{}ij\rangle_3$ denotes the third NN bonds. Here, we have introduced minus signs for each parameter, which is opposite to what is often used in the literature. For simplicity, we focus on a specific point in the parameter space, given by 
\begin{equation}
\begin{array}{lll}
{\rm FM}~K=1,& {\rm AFM}~\varGamma=-0.25,& {\rm AFM}~\varGamma'=-0.27,\\
{\rm FM}~J=0.37, & {\rm AFM}~J_3=-0.3, &
\end{array}\label{eq:parameters}
\end{equation}
which is close to a set of parameters identified in Ref.~\cite{maksimov2020rethinking} for $\alpha$-RuCl$_3$, resulting in a zigzag-ordered ground state obtained by DMRG simulations. Moreover, an in-plane magnetic field suppresses the long-ranged zigzag order and ultimately polarizes the state for sufficiently strong fields. Furthermore, it has been proposed that an intermediate KSL state may separate these phases.
To account for this effect, we introduce an additional Zeeman term in Eq.~\eqref{eq:HaRC3}. The total spin Hamiltonian is then given by
\begin{equation}
\mathcal{H}_{\rm S} = \mathcal{H}_{\rm S}^{\alpha-RC_3}+\sum_{i}\frac{h_{\text{in}}}{\sqrt{2}}\left(S^x_i-S^y_i\right),\label{eq:HaRC3h}
\end{equation}
where $h_{\text{in}}$ represents the in-plane magnetic field along the $[1\bar{1}0]$ direction. 

\begin{figure}
    \includegraphics[width=0.97\linewidth]{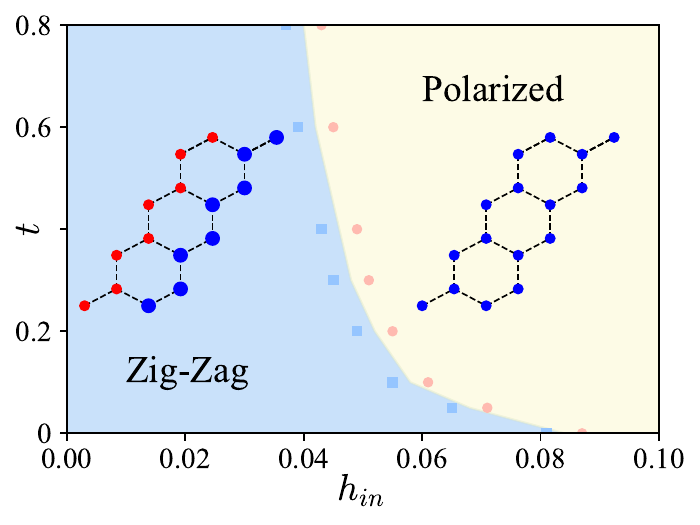}
    \caption{The ground-state phase diagram of hole-doped $K-J-\varGamma-\varGamma'-J_{3}$ model defined in Eq.~\eqref{eq:HaRC3h}; see also model parameters in Eq.~\eqref{eq:parameters}. The left inset shows a typical zigzag order at $(t,h_{in})=(0.3,0.03)$ and the right one shows a polarized state at $(t,h_{in})=(0.5,0.06)$. Positive and negative local magnetizations, $\langle S^x_j\rangle$, are represented by red and blue dots, respectively, with their radii scaled according to the magnitude of $\langle S^x_j\rangle$, with $\langle S^x_j\rangle\approx0.21$. The doping level is $\delta=0.0625$.}
    \label{fig:ZZ2P}
\end{figure}

We then investigate how hole doping modifies the phase boundary of the zigzag phase. We add the same kinetic energy term for the holes as in Eq.~\eqref{eq:H}, to the experimentally relevant Hamiltonian Eq.~\eqref{eq:HaRC3h} and compute the ground-state phase diagram with DMRG; see Fig.~\ref{fig:ZZ2P}. Remarkably, we find that the phase boundary between the zigzag and polarized phases is highly sensitive to hole doping. For a vanishing hopping strength, $t\rightarrow{}0$, the critical magnetic field required to polarize the spins is approximately $h^*_{in}\approx0.085$. However, already for a small hopping $t=0.8K$ and low hole doping of $\delta = 0.0625$, the critical magnetic field is reduced to $h^*_{in}\approx0.04$. A larger value of $t\approx1.8K$ can drive the zigzag order to a FM order (not shown). This can be understood by noting that $|t|\delta$ effectively acts as a FM Heisenberg exchange constant $J$. The polarized phase is distinct from the kinetic FM phase, as the latter develops magnetic order spontaneously. The significant effect induced by the kinetic energy of holes, therefore, is a complicating factor in identifying to correct Hamiltonian parameters in comparison to experimental measurements. 
\\


\noindent {\bf DISCUSSION}

\noindent
In summary, we have studied the hole-doped Kitaev-$\varGamma$ model using DMRG and effective parton descriptions. It is by now well known that the FM KSL is much more fragile with respect to the application of an external magnetic field compared to the AFM KSL~\cite{Zhu2018}. Similarly, our DMRG simulations suggest that the AFM KSL is also much more robust to hole doping. Meanwhile, distinct hole spectral functions have been numerically observed in FM and AFM KSLs, in which a dynamical Nagaoka ferromagnetism emerges in the FM KSL only~\cite{Wilhelm2023}. Concentrating on FM Kitaev exchange, we have shown here that small hole doping is sufficient to destabilize the KSL leading to a kinetic FM state generally coexisting with finite plaquette fluxes. 
We emphasize that, unlike the field-polarized phase in a magnetic field~\cite{janssen2019heisenberg}, the doping-induced FM order breaks the time-reversal symmetry spontaneously. Due to the particle-hole symmetry of our model, our conclusions also hold for charge doping.

We developed a parton mean-field theory, incorporating fermionic holons and bosonic magnons, which indeed shows that hole condensation gives rise to effective FM Heisenberg exchanges. 
We found that in the absence of off-diagonal exchange $\varGamma$, the spin polarization is along the [001] direction due to an order-by-disorder mechanism, which is also supported by DMRG simulations. A FM $\varGamma$ term switches the polarization direction to [111], which leads to a distinct topological FM phase with chiral boundary magnon excitations. 

Our work raises a whole range of questions for future research. First, one could investigate flux threading and topological entanglement of the FM phase coexisting with finite plaquette fluxes to probe the presence of topological order. Second, it would be interesting to search for other kinetic FM phases of Hubbard models hosting chiral magnon excitations. 
Third, to further investigate the transition between the KSL and kinetic FM phases, it would be desirable to develop a parton theory incorporating bosonic holons and {\it fermionic} spinons. In this context, such a theory could be the starting point for a self-consistent random phase approximation~\cite{Willsher2023} or Gutzwiller-boosted DMRG~\cite{Jin2021,Jin2022,Gabriel2021,Aghaei2020} in order to study the ordering instabilities and spin excitation spectra. 
Forth, in our real-time evolution results we could observe that the excitations of spin induce a (weak) response of charge, and vice versa. This could possibly be captured by a time-dependent parton mean-field theory in future studies to explore the interplay of charge and spin transport~\cite{minakawa2020majorana}.
Fifth, a study of doping effects for realistic model Hamiltonians could unveil a new mechanism (see Supplementary Note 4) for stabilizing the zigzag magnetic orders observed experimentally. This could also be important for refining the microscopic models of RuCl$_3$ and its peculiar magnetic field response, both of which remain poorly understood~\cite{maksimov2020rethinking}. 
Sixth, the origin of the observed thermal Hall effect of RuCl$_3$~\cite{kasahara2018majorana} is hotly debated~\cite{czajka2023planar} and it would be very interesting to explore how the topological magnons of the weakly doped Kitaev model studied here may contribute.

In conclusion, the Kitaev honeycomb model continues to be a fertile soil for novel physics. We expect the interplay of Kitaev spin exchange with kinetic electron motion to hold further surprises in the future. \\

\noindent {\bf DATA AVAILABILITY}

\noindent All data needed to evaluate the conclusions in the paper are present in the paper and/or the Supplementary Materials. \noindent The codes and data analysis are available on Zenodo upon reasonable request~\cite{Zenodo}. \\

\noindent {\bf ACKNOWLEDGEMENTS}

\noindent {\bf Acknowledgement} 

\noindent
We acknowledge support from the Imperial-TUM flagship partnership, the Deutsche Forschungsgemeinschaft (DFG, German Research Foundation) under Germany's Excellence Strategy--EXC--2111--390814868, DFG grants No. KN1254/1-2, KN1254/2-1, and TRR 360 - 492547816 and from the European Research Council (ERC) under the European Unions Horizon 2020 research and innovation programme (Grant Agreements No. 771537 and No. 851161), the International Centre for Theoretical Sciences (ICTS) for the program "Frustrated Metals and Insulators" (code: ICTS/frumi2022/9), as well as the Munich Quantum Valley, which is supported by the Bavarian state government with funds from the Hightech Agenda Bayern Plus. The numerical simulations in this work are based on the GraceQ project~\cite{graceq} and TeNPy Library~\cite{tenpy}.\\ 
\noindent {\bf Author contributions}  

\noindent H.-K.J. performed the numerical and analytical calculations and evaluated the data. The research was devised by M.K. and J.K. All authors contributed to analyzing the data, discussions, and the writing of the manuscript.\\
\noindent {\bf Competing interests} 

\noindent The authors declare that they have no competing interests.

\bibliography{dopeKitaev}

\clearpage

\begin{widetext}

\begin{large}
\begin{center}
 Supplemental Materials for \\ ``Kinetic Ferromagnetism and Topological Magnons of the Hole-Doped Kitaev Spin Liquid''
\end{center}
\end{large}

\hfill \break
\hfill \break

This Supplemental Material includes more details of the DMRG calculations and parton theory. 
	In Supplementary Note 1, we show the considered honeycomb lattice and the corresponding one-dimensional path for DMRG simulations. 
	In Supplementary Note 2, we show additional information about the DMRG results. 
	In Supplementary Note 3, we show details of the parton mean-field theory.
	In Supplementary Note 4, we provide additional results for the zig-zag order.
	In Supplementary Note 5, we provide more information about the spin-1/2 moment around the exactly  solvable point of  $t=0$.
	
	\section*{Supplementary Note 1: Honeycomb lattice and one-dimensional path}\label{app:1d}
	
	\begin{figure}[!h]
		\includegraphics[width=0.5\linewidth]{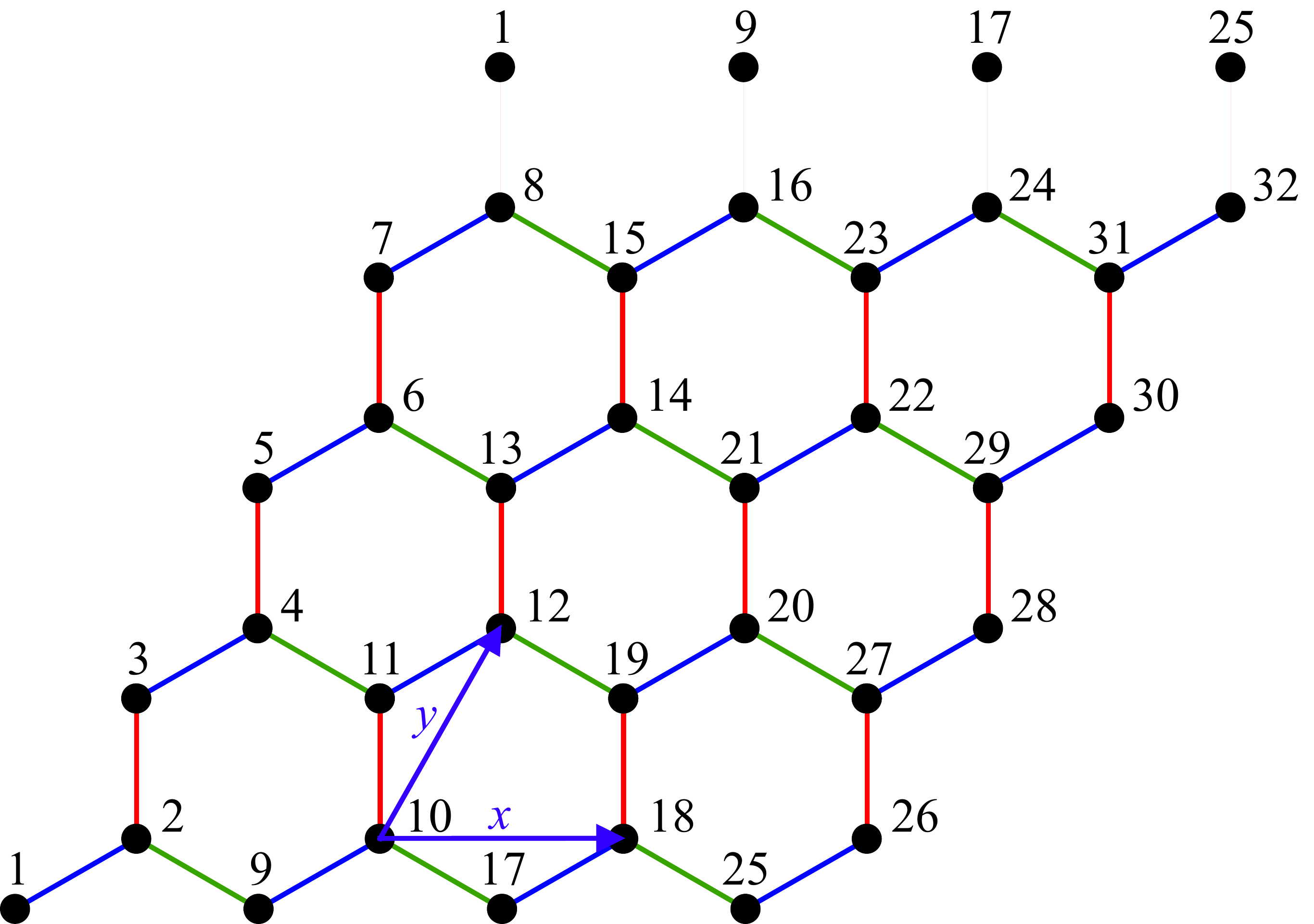}
		\caption{Schematics of the labeling scheme for a honeycomb lattice as a cylindrical geometry with primitive vectors $\vec{x}$ and $\vec{y}$. The blue, green, and red lines represent the $x$-, $y$-, and $z$-type bonds for a Kitaev honeycomb model, respectively. The periodic boundary condition has been imposed along the $y$ direction, as indicated by the red dashed lines.}\label{fig:1dpath}
	\end{figure}

	To employ the MPS-based methods, one must establish a site ordering for the honeycomb lattice. 
	This can be achieved by assigning an integer from 1 to $2\mathrm{N}_c$ to each lattice site, where $\mathrm{N}_c$ is the number of unit cells. As shown in Fig.~\ref{fig:1dpath}, we utilize a site-labeling scheme for the honeycomb lattice on a YC4 ($L_y=4$) cylinder with length $L_x$ with primitive vectors $\vec{x}$ and $\vec{y}$. Note that the $x$-, $y$-, and $z$-type bonds for a Kitaev honeycomb model are marked by the blue, green, and red lines, respectively.

	\section*{Supplementary Note 2: Additional DMRG Data}\label{app:ndata}

	\subsection*{Ground-state calculations}
	
	For the DMRG simulations we make explicit use of a U(1) quantum number to control the level of hole doping. The bond dimension of DMRG is kept as large as $\chi=4000$, resulting in a typical truncation error of $\approx 7\times{}10^{-7}$.  
	\begin{figure}
		\includegraphics[width=0.95\linewidth]{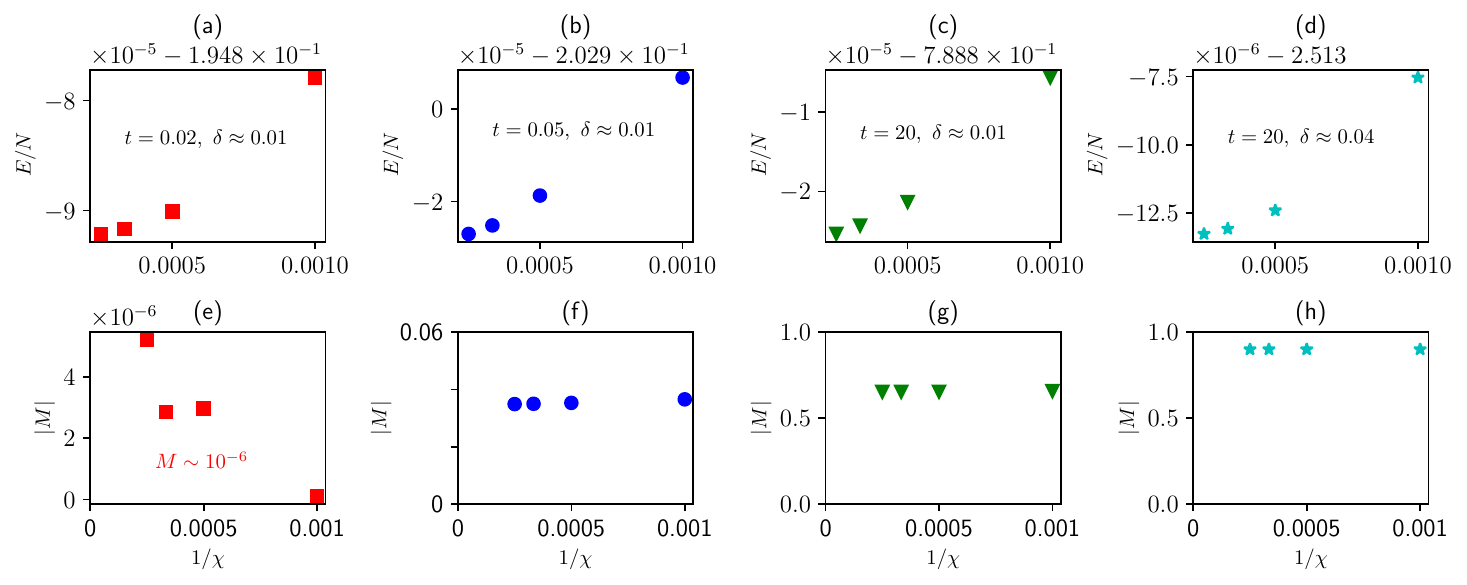}
		\caption{(a-d) Ground-state energies per site as functions of inverse MPS bond dimension $1/\chi$ for several typical model parameters. (e-h): Magnetizations as functions of  $1/\chi$. The calculations are performed on YC4 cylinders with $L_x=12$ and $\varGamma=0$.
		}\label{fig:scaling}
	\end{figure}
	To analyze the convergence of our DMRG simulations, we study the scaling behavior of the ground-state energy and magnetization with the inverse DMRG bond dimension $1/\chi$.
	As shown in Fig.~\ref{fig:scaling},  the ground-state energy varies weakly with $\chi$ for $\chi>1000$, suggesting a faithful convergence in our DMRG simulations. In the slow-hole regime (Kitaev spin liquid phase), the magnetization fluctuates around the magnitude of truncation error, indicating the absence of magnetic order. In the fast-hole regime (FM-ordered phase),  the  magnetization extrapolates to a finite value when $1/\chi \rightarrow 0$, indicating that the kinetic FM order is stable when increasing the bond dimension.
	
	\begin{figure}
		\includegraphics[width=0.55\linewidth]{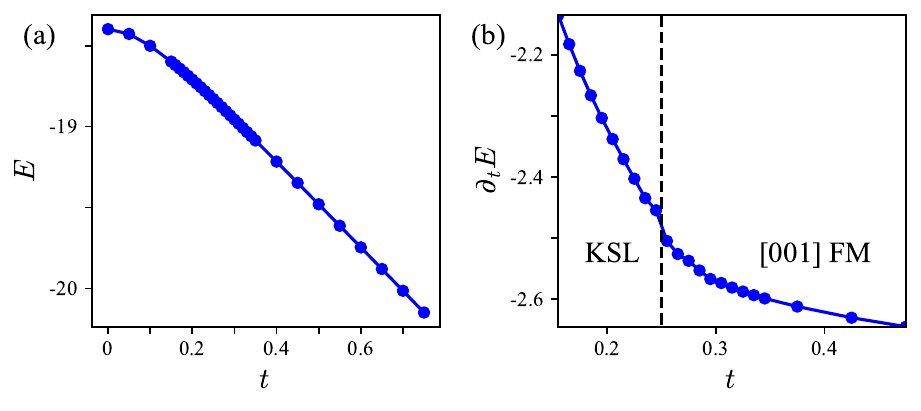}
		\caption{The ground state energy (a) and its first-order derivative (b) for the $t$-$K$ model as a function of $t$ on a single-hole-doped YC4 cylinder with $\delta\approx0.01$.}\label{fig:EdE}
	\end{figure}

	Fig.~\ref{fig:EdE} displays the ground-state energy $E$ and its first-order derivative $\partial_tE$ as functions of hopping strength $t$. Note that $\partial_tE$ in Fig.~\ref{fig:EdE}(b) exhibits a kink at $t^*\approx{}0.28{}K$ which suggests that the phase transition is of the second order.
	
	\subsection{Superconducting correlation functions}
	
	\begin{figure}
		\includegraphics[width=1\linewidth]{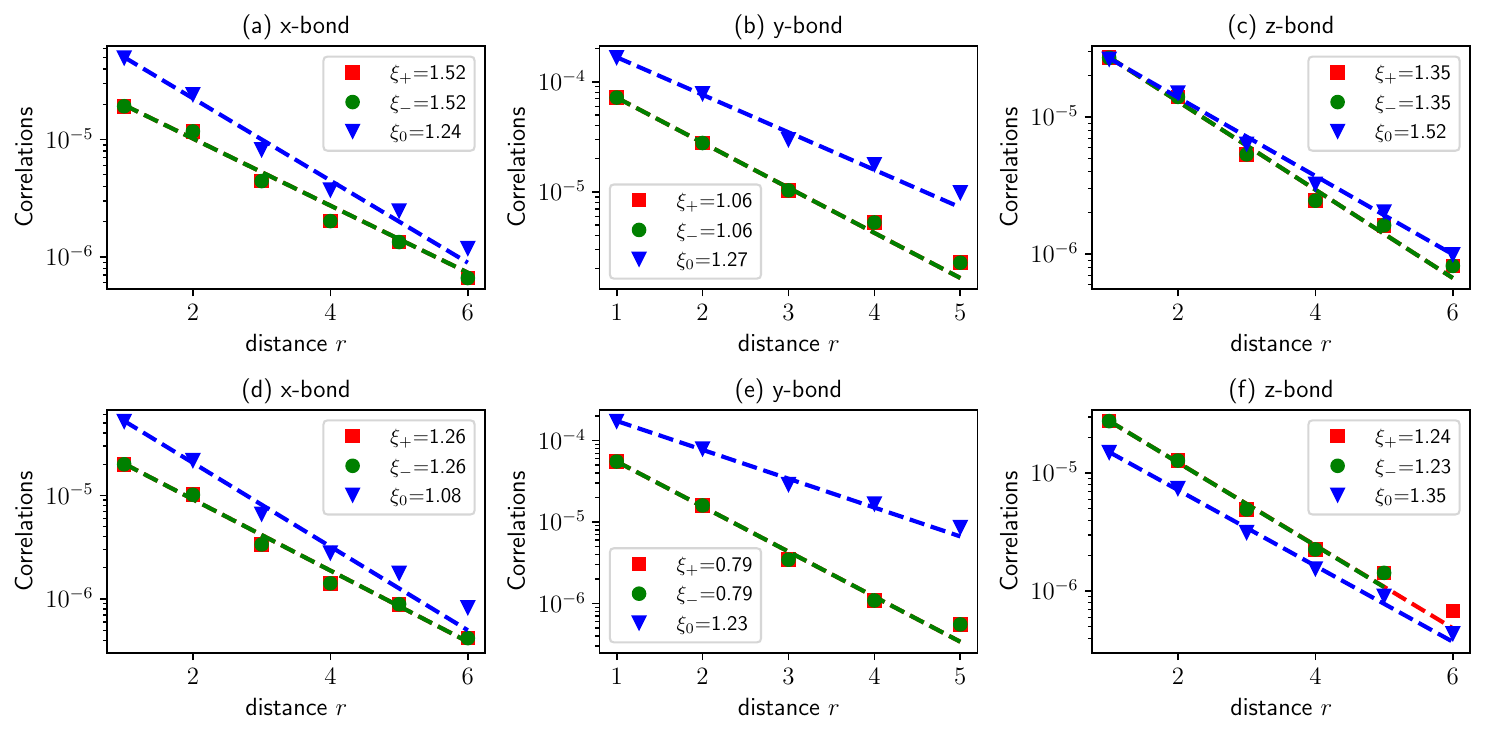}
		\caption{The odd-parity pairing-pairing correlation function (semi-log plot) [see Eqs.~\eqref{eq:pair1}-\eqref{eq:pair3}]  obtained on a YC4 cylinder with $\varGamma=0$, $\delta\approx0.04$, and $\chi=4000$. For (a-c) we set $t=20K$ and for (d-e) $t=2K$. The even-parity correlation functions, which are not shown here, are significantly smaller than the odd-parity functions ($\Phi_{e}\sim{}10^{-12}$).	}\label{fig:figS2}
	\end{figure}
	
	The multiple-hole doping allows us to analyze the equal-time pairing-pairing correlation function. 
	A diagnostic of the superconducting order is the  pair–pair correlation function, defined as
	\begin{equation}
		\Phi_{a}(r)=\langle{}\Delta^\dag_{a}(r_0)\Delta_{a}(r_0+r)\rangle,\label{eq:pair1}
	\end{equation} 
	where $\Delta^\dag_{a}(r_0)$ is SC pair-field creation operator with pairing index $a$  defined on the $\gamma$-type ($\gamma=x,y,z$) NN bond $r_0$ and  $r$ is the distance along the cylinder (the $\hat{x}$ direction). We consider one even-parity pairing as
	\begin{equation}
		\Delta^\dag_e(r_0)=\frac{1}{\sqrt{2}}\left[c^\dag_{i_0,\uparrow}c^\dag_{j_0,\downarrow}-c^\dag_{i_0,\downarrow}c^\dag_{j_0,\uparrow}\right],,\label{eq:pair2}
	\end{equation} 
	where $i_0$ and $j_0$ denote two lattice sites on the NN bond $r_0$.  Similarly, we define three odd-parity pairing functions as
	\begin{equation}
		\begin{split}
			&\Delta^\dag_+(r_0)=c^\dag_{i_0,\uparrow}c^\dag_{j_0,\uparrow},	\\
			&\Delta^\dag_-(r_0)=c^\dag_{i_0,\downarrow}c^\dag_{j_0,\downarrow},	\\
			&\Delta^\dag_0(r_0)=\frac{1}{\sqrt{2}}\left[c^\dag_{i_0,\uparrow}c^\dag_{j_0,\downarrow}+c^\dag_{i_0,\downarrow}c^\dag_{j_0,\uparrow}\right].	\\
		\end{split},\label{eq:pair3}
	\end{equation} 
	
	In practice, we set  $i_0$ and $j_0$ being the reference bond at $x\sim{}L_x/4$ to minimize the boundary effect. We study the above-mentioned pair–pair correlation functions on the $x$-, $y$-, and $z$-type bonds for $\delta\approx0.04$ and as shown in Fig.~\ref{fig:figS2}, we find that all of them exhibit exponential decays, 
	$$\Phi_{a}\sim{}e^{-r/\xi_a}.$$  
	The correlation lengths $\chi_a$ for different pairing types are summarized in Fig.~\ref{fig:figS2}.
	This observation suggests the absence of superconductivity in the weak doping limit.

	\subsection*{Real-time dynamics}
	
	The real-time dynamics is performed by the standard MPO-evolution algorithm~\cite{Zaletel2015}. A small time step d$t=0.01[K]^{-1}$ has been used and  At each intermediate step of the real-time evolution, one needs to truncate the MPO-evolved MPS. In order to estimate the accuracy of the final MPS, we introduce the accumulated truncation error defined by
	\begin{align}
		\epsilon_{\mathrm{trunc}}(D)=\sum_{j}\epsilon_{j}(D)
		\label{eq:trunc_error}
	\end{align}
	where $\epsilon_{j}(D)$ is the sum of the discarded squared singular values at the $j$-th bond. Meanwhile, we also track the growth of entanglement entropy (EE) of the evolved MPS. As shown in Fig.~\ref{fig:eete}, the truncation errors are close to zero when $t<3$ and are always less than $10^{-5}$ during the whole evolution period of $5[K]^{-1}$. Note that EE displays a jump at $t=0$ because we introduce a magnon-like excitation at the boundary. Then, the EE gradually increases until it saturates to a plateau value. 
	
	\begin{figure}[!b]
		\includegraphics[width=0.5\linewidth]{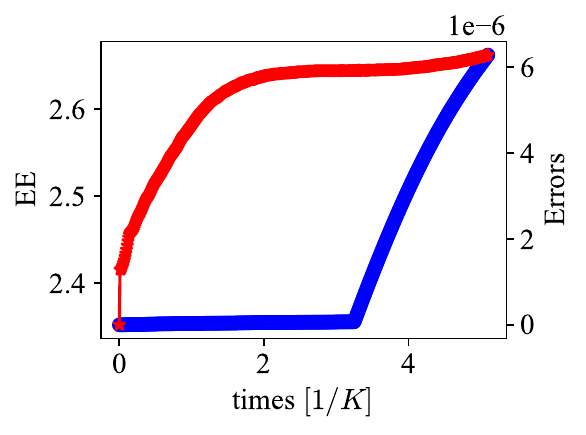}
		\caption{The evolutions of entanglement entropy (EE, left axis) and truncation errors (right axis) during the real-time evolution. The simulations are performed on a YC3 cylinder with $\varGamma=0.05K$, $t=10K$, and $\delta\approx0.013$ (single hole doped).}\label{fig:eete}
	\end{figure}

	\section*{Supplementary Note 3: Parton mean-field theory}\label{app:pmft}
	In this section, we provide a detailed description of the parton mean-field theory defined in Eq.~(4) of the main text. We summarize the main results as follows: We decouple the $t$-$K$-$\varGamma$ model in the FM phase into separate hole and spin parts, $$\mathcal{H}\approx{}\mathcal{H}_{\rm h}+\mathcal{H}_{\rm s},$$ using a mean-field theory. The hole part $\mathcal{H}_{\rm h}$ can be effectively described by $\mathcal{H}_{\rm h}$ in Eq.~\eqref{eq:Hh}. For the spin part, the ground-state energy corrected by the zero-point quantum fluctuation is $E_s$ in Eq.~\eqref{eq:Es2}, which leads to a quantum order-by-disorder phenomenon for $\varGamma=0$.  The corresponding magnon excitations can be effectively described by a spin-wave theory with Hamiltonian $\mathcal{H}_{\rm s}\approx{}H^{(2)}_s+\delta H^{(2)}_s$, as defined in Eqs.~\eqref{eq:Hs2} and \eqref{eq:Hs24}, respectively.
	
	As mentioned in the main text, we represent fermionic electron operators in terms of fermionic holons $h^{}_j$ and bosonic spinons $b^{}_{j,s}$:
	\begin{eqnarray}
		&c^\dag_{j,s}=h^{}_j{}b^\dag_{j,s},\qquad c^{}_{j,s}=h_j^\dag{}b^{}_{j,s}.\label{eq:cfermion}
	\end{eqnarray}
	This representation, which is particularly suitable for studying symmetry-breaking states, enlarges the local Hilbert space. In order to restore the original Hilbert space, one needs to impose the local constraints, $h^\dag_jh_j=\delta{}$ and $\sum_{s}b^\dag_{j,s}b_{j,s}=1-\delta{}$, for every lattice site $j$. By means of Eq.~\eqref{eq:cfermion}, the kinetic terms can be rewritten as 
	\begin{equation}
		\sum_{s}\mathcal{P}_{g}c^\dagger_{j,s}c^{}_{k,s}\mathcal{P}_{g}=-\sum_{s}h^\dag_kh^{}_jb^\dag_{j,s}b^{}_{k,s}.
	\end{equation}
	We introduce a vector of spinons, $\mathbf{b}^\dag_j = \left(b^\dag_{j,\uparrow},b^\dag_{j,\downarrow}\right)$, and express the bilinear spin-spin interactions in a compact form as
	\begin{eqnarray}
		&&S^{\alpha}_jS^{\beta}_j=h^{}_jh^\dag_jh^{}_kh^\dag_k\hat{S}^\alpha_{j}\hat{S}^\beta_{k},\\
		&&\hat{S}^\alpha_{j}\hat{S}^\beta_{k}=\frac{1}{4}\left(\mathbf{b}^\dag_j\sigma^\alpha\mathbf{b}^{}_j\right)\left(\mathbf{b}^\dag_k\sigma^\beta\mathbf{b}^{}_k\right).
	\end{eqnarray}
	Then, we can obtain the spin part Hamiltonian on the NN bonds $\langle{jk}\rangle\in\gamma$, as 
	$$\hat{H}^\gamma_{jk}=-K\hat{S}^\gamma_j\hat{S}^\gamma_k-\varGamma\left(\hat{S}^\alpha_j\hat{S}^\beta_k+\hat{S}^\beta_j\hat{S}^\alpha_k\right).$$
	
	Assuming a translational invariant ansatz,  the Hamiltonians for the hole and spin parts can be further decoupled into several quadratic terms by introducing mean-field parameters on the NN bonds $\langle{jk}\rangle\in\gamma$, which should be determined self-consistently from
	\begin{eqnarray}
		&&P_{\gamma}=\langle{}\hat{P}_{jk}\rangle,\textrm{~~with~~} \hat{P}_{jk} = h^\dag_{j}h^{}_{k}, \label{eq:P}\\
		&&F^{\gamma}_{s_1,s_2}=\langle{}\hat{F}_{js_1,ks_2} \rangle,\textrm{~~with~~} \hat{F}_{js_1,ks_2} = b^\dag_{j,s_1}b^{}_{k,s2},\label{eq:F}\\
		&&A^{\gamma}_{s_1,s_2}=\langle{}\hat{A}_{js_1,ks_2} \rangle,\textrm{~~with~~}\hat{A}_{js_1,ks_2} = b^{}_{j,s_1}b^{}_{k,s2}. \label{eq:A}
	\end{eqnarray}
	Here $P_{\gamma}$ represents the holon mobility. And generally, $F^{\gamma}$ and $A^{\gamma}$ represent the ferromagnetic and antiferromagnetic short-range correlations, respectively. 
	
	After implementing the following decoupling scheme as
	$$
	(1-h^\dag_j{}h_j)(1-h^\dag_k{}h_k)\rightarrow{}-P_{\gamma}h^\dag_{j}h^{}_{k}-P_{\gamma}h^\dag_{k}h^{}_{j}+P_{\gamma}P^*_{\gamma},
	$$	
	the mean-field Hamiltonian for the hole part reads
	\begin{equation}
		\begin{split}
			\mathcal{H}_{\rm h}=&\sum_{\langle{jk}\rangle\in\gamma,s}-t\left(F^{\gamma}_{ss}h^\dag_{k}h^{}_{j}+{\rm h.c.}\right)+\sum_{j}\mu{}h^\dag_{j}h^{}_{j}+\\&\sum_{\langle{jk}\rangle}\langle\hat{H}^\gamma_{jk}\rangle\left(-P^*_{\gamma}h^\dag_{j}h^{}_{k}-P_{\gamma}h^\dag_{k}h^{}_{j}+P_{\gamma}P^*_{\gamma}\right),\label{eq:Hh}
		\end{split}
	\end{equation}
	where $\mu$ is the Lagrange multiplier to tune the filling number of holons such that $\langle{}h^\dag_{j}h^{}_{j}\rangle=\delta$. Since $P_\gamma\sim{}\delta$, the terms in the last line in  Eq.~\eqref{eq:Hh} can be neglected in the small doping limit.
	
	The four-boson terms $\hat{H}^\gamma_{jk}$ in the spin part Hamiltonian need to be further decoupled in order to achieve a quadratic form. For instance, we employ the following mean-field treatment for $\hat{H}^x_{jk}$, which enables us to decouple the Kitaev-type interactions as follows:
	\begin{eqnarray*}
		&-&\sum_{s_1s_2}\frac{K}{4}\left[F^x_{s_1,\bar{s}_2}\hat{F}_{ks_2,j\bar{s}_1}+F^{x,*}_{\bar{s}_1,s_2}\hat{F}_{js_1,k\bar{s}_2}-F^x_{s_1,\bar{s}_2}F^{x,*}_{\bar{s}_{1},s_2}\right]\\
		&-&\sum_{s_1s_2}\frac{K}{4}\left[A^{x,*}_{s_1,s_2}\hat{A}_{j\bar{s}_1,k\bar{s}_2}+A^x_{\bar{s}_1,\bar{s}_2}A^{\dag}_{js_1,ks_2}-A^{x,*}_{s_1,s_2}A^x_{\bar{s}_1,\bar{s}_2}\right],
	\end{eqnarray*}
	and the anisotropic spin-spin interaction as
	\begin{eqnarray*}		&i&\sum_{s_1s_2}(-1)^{s_1+s_2}\frac{\varGamma}{4}\left[F^{x}_{s_1,s_2}\hat{F}_{ks_2,j\bar{s}_1}+F^{x,*}_{\bar{s}_1,s_2}\hat{F}_{js_1,ks_2}-F^{x}_{s_1,s_2}F^{x,*}_{\bar{s}_1,s_2}\right]\\
		+&i&\sum_{s_1s_2}(-1)^{s_1+s_2}\frac{\varGamma}{4}\left[F^{x}_{s_1,\bar{s}_2}\hat{F}_{ks_2,js_1}+F^{x,*}_{s_1,s_2}\hat{F}_{js_1,k\bar{s}_2}-F^{x}_{s_1,\bar{s}_2}F^{x,*}_{s_1,s_2}\right]\\
		+&i&\sum_{s_1s_2}(-1)^{s_1+s_2}\frac{\varGamma}{4}\left[A^{x,*}_{s_1,s_2}\hat{A}_{j\bar{s}_1,ks_2}+A^x_{\bar{s}_1,s_2}\hat{A}^{\dag}_{js_1,ks_2}-A^{x,*}_{s_1,s_2}A^x_{\bar{s}_1,s_2}\right]\\
		+&i&\sum_{s_1s_2}(-1)^{s_1+s_2}\frac{\varGamma}{4}\left[A^{x,*}_{s_1,s_2}\hat{A}_{js_1,k\bar{s}_2}+A^x_{s_1,\bar{s}_2}\hat{A}^{\dag}_{js_1,ks_2}-A^{x,*}_{s_1,s_2}A^x_{s_1,\bar{s}_2}\right].\\
	\end{eqnarray*}
	Similar decomposition schemes are also applied for $\hat{H}^y_{jk}$ and $\hat{H}^z_{jk}$.
	For convenience, we further define
	\begin{equation}
		D_\gamma=\langle(1-h^\dag_jh^{}_j)(1-h^\dag_kh^{}_k)\rangle\label{eq:Delta}
	\end{equation}
	on a  NN bond $\langle{jk}\rangle\in\gamma$.  Since the system exhibits translational invariance, we can perform a Fourier transform on the bosonic spinons and introduce a spinon vector at momentum $\mathbf{k}$ as
	\begin{equation}
		\begin{array}{ccccc}
			\psi^\dagger_{\bf k}=&\Big{(}b^\dagger_{{\bf k},0,\uparrow}&b^\dagger_{{\bf k},1,\uparrow}&b^\dagger_{{\bf k},0,\downarrow}&b^\dagger_{{\bf k},1,\downarrow}\Big{)},
		\end{array}
	\end{equation}
	where the additional index $l=0,1$ denotes the two sublattices of a honeycomb lattice.
	Using $D_\gamma$, $\psi^\dagger_{\bf k}$, and a Lagrange multiplier $\lambda$ to tune the condition such that $n_b\equiv{}\langle{}b^\dag_{j,s}b^{}_{j,s}\rangle{}=1-\delta$,  we arrive at the mean-field Hamiltonian for the spinon part
	\begin{equation}
		\begin{split}
			\mathcal{H}_{\rm s}=&t\sum_{\langle{jk}\rangle\in\gamma,s}\left(P^*_\gamma{}b^\dag_{js}b^{}_{ks}+{\rm h.c.}\right)+\sum_{j,s}\lambda{}\left(b^\dag_{j,s}b^{}_{j,s}-1+\delta\right)+\\
			&\frac{1}{4}\sum_{\bf k}\sum_{\gamma=x,y,z}D_\gamma
			\left(\psi^\dag_{\bf k},\psi^{}_{\bf -k}\right)
			\left(\begin{array}{cc}
				T^\gamma_{\bf k} & A^\gamma_{\bf k}\\
				(A^\gamma_{\bf k})^\dag & (T^\gamma_{\bf -k})^T
			\end{array}\right)
			\left(\begin{array}{c}
				\psi^{}_{\bf k} \\
				\psi^\dag_{\bf -k}
			\end{array}\right),\\
		\end{split}\label{eq:sbmft}
	\end{equation}
	where two $4\times{}4$ matrices $T^\gamma_{\mathbf{k}}$ and $D^\gamma_{\mathbf{k}}$ correspond to bosonic hopping and pairing terms.
	
	The large-$N$ limit of $\mathcal{H}_{\rm s}$ is taken with a parameter~\cite{Sachdev1992}  
	$$\kappa\equiv{}n_b/N$$ 
	being fixed. For a small value of $\kappa$, the Schwinger-boson mean-field theory may give rise to a gapful QSL, while a magnetically ordered state with bosonic spinon condensation appears at large $\kappa$. In our study, we focus on its realistic limit with $\kappa=n_b=1-\delta$, i.e., $N=1$. Here the explicit forms of the terms in  Eq.~\eqref{eq:sbmft} are
	\begin{small}
		\begin{flalign*}
			&\quad{}T^x_{\bf k}=-\frac{K}{2}
			\left(\begin{array}{ccccc}
				& F^{x,*}_{\downarrow\downarrow} & & F^{x,*}_{\downarrow\uparrow}\\
				F^{x}_{\downarrow\downarrow} & & F^{x}_{\uparrow\downarrow} &\\
				& F^{x,*}_{\uparrow\downarrow} & & F^{x,*}_{\uparrow\uparrow}\\
				F^{x}_{\downarrow\uparrow} & & F^{x}_{\uparrow\uparrow} &\\
			\end{array}\right)+\frac{i\varGamma}{2}
			\left(\begin{array}{ccccc}
				& F^{x,*}_{\downarrow\uparrow} -F^{x,*}_{\uparrow\downarrow}& & -F^{x,*}_{\downarrow\downarrow}+F^{x,*}_{\uparrow\uparrow}\\
				-F^{x}_{\downarrow\uparrow}+F^{x}_{\uparrow\downarrow} & & F^{x}_{\uparrow\uparrow}-F^{x}_{\downarrow\downarrow} &\\
				& -F^{x,*}_{\uparrow\uparrow}+F^{x,*}_{\downarrow\downarrow} & & F^{x,*}_{\uparrow\downarrow}-F^{x,*}_{\downarrow\uparrow}\\
				F^{x}_{\downarrow\downarrow}-F^{x}_{\uparrow\uparrow} & & -F^{x}_{\uparrow\downarrow}+F^{x}_{\downarrow\uparrow} &\\
			\end{array}\right),&
		\end{flalign*}
		\begin{flalign*}
			&\quad{}T^y_{\bf k}=-\frac{K}{2}
			\left(\begin{array}{ccccc}
				& F^{y,*}_{\downarrow\downarrow}e^{ik_x} & & -F^{y,*}_{\downarrow\uparrow}e^{ik_x}\\
				F^{y}_{\downarrow\downarrow}e^{-ik_x} & & -F^{y}_{\uparrow\downarrow}e^{-ik_x} &\\
				& -F^{y,*}_{\uparrow\downarrow}e^{ik_x} & & F^{y,*}_{\uparrow\uparrow}e^{ik_x}\\
				-F^{y}_{\downarrow\uparrow}e^{-ik_x} & & F^{y}_{\uparrow\uparrow}e^{-ik_x}&\\
			\end{array}\right)-\frac{\varGamma}{2}
			\left(\begin{array}{ccccc}
				& (F^{y,*}_{\uparrow\downarrow}+F^{y,*}_{\downarrow\uparrow})e^{ik_x} & & (F^{y,*}_{\uparrow\uparrow}-F^{y,*}_{\downarrow\downarrow})e^{ik_x}\\
				(F^{y}_{\uparrow\downarrow}+F^{y}_{\downarrow\uparrow})e^{-ik_x} & & (F^{y}_{\uparrow\uparrow}-F^{y}_{\downarrow\downarrow})e^{-ik_x} &\\
				& (F^{y,*}_{\uparrow\uparrow}-F^{y,*}_{\downarrow\downarrow})e^{ik_x} & & -(F^{y,*}_{\downarrow\uparrow}+F^{y,*}_{\uparrow\downarrow})e^{ik_x}\\
				(F^{y}_{\uparrow\uparrow}-F^{y}_{\downarrow\downarrow})e^{-ik_x} & & -(F^{y}_{\downarrow\uparrow}+F^{y}_{\uparrow\downarrow})e^{-ik_x}&\\
			\end{array}\right),&
		\end{flalign*}
		\begin{flalign*}
			&\quad{}T^z_{\bf k}=-\frac{K}{2}
			\left(\begin{array}{ccccc}
				& F^{z,*}_{\uparrow\uparrow}e^{ik_y} & & -F^{z,*}_{\uparrow\downarrow}e^{ik_y}\\
				F^{z}_{\uparrow\uparrow}e^{-ik_y} & & -F^{z}_{\downarrow\uparrow}e^{-ik_y} &\\
				& -F^{z,*}_{\downarrow\uparrow}e^{ik_y} & & F^{z,*}_{\downarrow\downarrow}e^{ik_y}\\
				-F^{z}_{\uparrow\downarrow}e^{-ik_y} & & F^{z}_{\downarrow\downarrow}e^{-ik_y}&\\
			\end{array}\right)+\frac{i\varGamma}{2}
			\left(\begin{array}{ccccc}
				& 0 & & 2F^{z,*}_{\downarrow\uparrow}e^{ik_y}\\
				0 & & 2F^{z}_{\uparrow\downarrow}e^{-ik_y} &\\
				& -2F^{z,*}_{\uparrow\downarrow}e^{ik_y} & & 0\\
				-2F^{z}_{\downarrow\uparrow}e^{-ik_y} & & 0&\\
			\end{array}\right),&
		\end{flalign*}
		\begin{flalign*}
			&\quad{}D^x_{\bf k}=-\frac{K}{2}\left(\begin{array}{ccccc}
				& A^{x}_{\downarrow\downarrow} & & A^{x}_{\downarrow\uparrow}\\
				A^{x}_{\downarrow\downarrow} & & A^{x}_{\uparrow\downarrow} &\\
				& A^{x}_{\uparrow\downarrow} & & A^{x}_{\uparrow\uparrow}\\
				A^{x}_{\downarrow\uparrow} & & A^{x}_{\uparrow\uparrow} &\\
			\end{array}\right) 
			+\frac{i\varGamma}{2}
			\left(\begin{array}{ccccc}
				& A^{x}_{\downarrow\uparrow}+A^{x}_{\uparrow\downarrow} & & -A^{x}_{\downarrow\downarrow}-A^{x}_{\uparrow\uparrow}\\
				A^{x}_{\downarrow\uparrow}+A^{x}_{\uparrow\downarrow} & & -A^{x}_{\downarrow\downarrow}-A^{x}_{\uparrow\uparrow} &\\
				& -A^{x}_{\downarrow\downarrow}-A^{x}_{\uparrow\uparrow} & & A^{x}_{\downarrow\uparrow}+A^{x}_{\uparrow\downarrow}\\
				-A^{x}_{\downarrow\downarrow}-A^{x}_{\uparrow\uparrow} & & A^{x}_{\downarrow\uparrow}+A^{x}_{\uparrow\downarrow} &\\
			\end{array}\right),&
		\end{flalign*}
		\begin{flalign*}
			&\quad{}D^y_{\bf k}=-\frac{K}{2}
			\left(\begin{array}{ccccc}
				& -A^{y}_{\downarrow\downarrow}e^{ik_x} & & A^{y}_{\downarrow\uparrow}e^{ik_x}\\
				-A^{y}_{\downarrow\downarrow}e^{-ik_x} & & A^{y}_{\uparrow\downarrow}e^{-ik_x} &\\
				& A^{y}_{\uparrow\downarrow}e^{ik_x} & & -A^{y}_{\uparrow\uparrow}e^{ik_x}\\
				A^{y}_{\downarrow\uparrow}e^{-ik_x} & & -A^{y}_{\uparrow\uparrow}e^{-ik_x} &\\
			\end{array}\right)-\frac{\varGamma}{2}
			\left(\begin{array}{ccccc}
				& (A^{y}_{\downarrow\uparrow}+A^{y}_{\uparrow\downarrow})e^{ik_x} & & (A^{y}_{\uparrow\uparrow}-A^{y}_{\downarrow\downarrow})e^{ik_x}\\
				(A^{y}_{\downarrow\uparrow}+A^{y}_{\uparrow\downarrow})e^{-ik_x} & & (A^{y}_{\uparrow\uparrow}-A^{y}_{\downarrow\downarrow})e^{-ik_x} &\\
				& (A^{y}_{\uparrow\uparrow}-A^{y}_{\downarrow\downarrow})e^{ik_x} & & -(A^{y}_{\downarrow\uparrow}+A^{y}_{\uparrow\downarrow})e^{ik_x}\\
				(A^{y}_{\uparrow\uparrow}-A^{y}_{\downarrow\downarrow})e^{-ik_x} & & -(A^{y}_{\downarrow\uparrow}+A^{y}_{\uparrow\downarrow})e^{-ik_x} &\\
			\end{array}\right),&
		\end{flalign*}
		\begin{flalign*}
			&\quad{}D^z_{\bf k}=-\frac{K}{2}	
			\left(\begin{array}{ccccc}
				& A^{z}_{\uparrow\uparrow}e^{ik_y} & & -A^{z}_{\uparrow\downarrow}e^{ik_y}\\
				A^{z}_{\uparrow\uparrow}e^{-ik_y} & & -A^{z}_{\downarrow\uparrow}e^{-ik_y} &\\
				& -A^{z}_{\downarrow\uparrow}e^{ik_y} & & A^{y}_{\downarrow\downarrow}e^{ik_y}\\
				-A^{y}_{\uparrow\downarrow}e^{-ik_y} & & A^{y}_{\downarrow\downarrow}e^{-ik_y} &\\
			\end{array}\right)+\frac{i\varGamma}{2}
			\left(\begin{array}{ccccc}
				& 2A^{z}_{\downarrow\downarrow}e^{ik_y} & & \\
				2A^{z}_{\downarrow\downarrow}e^{-ik_y} & &  &\\
				&  & & -2A^{z}_{\uparrow\uparrow}e^{ik_y}\\
				& & -2A^{z}_{\uparrow\uparrow}e^{-ik_y} &\\
			\end{array}\right).&
		\end{flalign*}
	\end{small}
	
	The quadratic Hamiltonians $\mathcal{H}_{\rm h}$ and $\mathcal{H}_{\rm s}$ can be diagonalized by eigenvalue decomposition and bosonic Bogolyubov transformation, respectively. Here we denote the dispersion relations for the Schwinger bosons (spinons) as $\epsilon_{{\bf k},n}$ with $n=1,2,3,4$. This enables us to numerically obtain the self-consistent solution for Eqs.~\eqref{eq:P}-\eqref{eq:A} through an iterative scheme. For instance, single occupancy leads to 
	\begin{equation}
		N = \langle\sum_{j,s}b^\dag_{j,s}b^{}_{j,s}\rangle = \rho+\frac{1}{2\mathrm{N}_c}\sum_{\epsilon_{\bf k}>0}\sum_{mn}[V^{12}_{\bf{k}}]_{mn}[V^{12}_{\bf{k}}]^\dagger_{mn},
	\end{equation} 
	where $\rho$ denote the condensate density, and $4\times{}4$ matrices $V^{11}_{\bf{k}}$, $V^{12}_{\bf{k}}$, $V^{21}_{\bf{k}}$, and $V^{22}_{\bf{k}}$ form the symplectic matrix $M_{\bf k}$ diagonalizing the Hamiltonian in Eq.~\eqref{eq:sbmft}, as $	M_{\bf k}
	=\left(\begin{smallmatrix}
		V^{11}_{\bf{k}} & V^{12}_{\bf{k}}\\
		V^{21}_{\bf{k}} & V^{22}_{\bf{k}}
	\end{smallmatrix}\right)$. 
	When $\rho>0$ is a finite value, the Schwinger bosons condense at specific momenta $\bf{k}^{\ast}$ with zero-energy modes $\epsilon_{{\bf k}^*,n}=0$, namely, $\langle{}b^\dag_{\bf{}k^*}\rangle=(\langle{}b^{}_{\bf{}k^*}\rangle)^*\sim{}\sqrt{\rho}$. This leads to a long-range magnetic ordered state. For instance, a ferromagnetic order along the $z$ direction means $\langle{}b^\dag_{j,\uparrow}\rangle=\langle{}b^{}_{j,\uparrow}\rangle=\sqrt{\rho}=\sqrt{\kappa}$ and similarly, one along the $x$ direction means $\langle{}b^\dag_{j,\uparrow}\rangle=\langle{}b^{}_{j,\uparrow}\rangle=\langle{}b^{\dag}_{j,\downarrow}\rangle=\langle{}b^{}_{j,\downarrow}\rangle=\sqrt{\kappa/2}$. 
	When $\rho=0$, all eigenmodes of Eq.~\eqref{eq:sbmft} are gapped and the corresponding ground state is a disordered quantum spin liquid state. When $A^\gamma=0$ and thereby $V^{12}_{\bf k}=0$, one must have $\rho=\kappa=1-\delta$ corresponding to a ferromagnetic ordered ground state.
	
	When carrying out the calculations, different randomly generated mean-field parameters are selected to initialize the iterations. However, after careful investigations, we conclude that, in the parameter regime of FM Kitaev coupling and subdominant $\varGamma$, there is no gapped QSL solution. It is well-known from the exact solution that the Kitaev honeycomb model exhibits a gapless $Z_2$ QSL. Hence, it is not surprising that the Schwinger-boson mean-field theory fails to describe the QSL phase for a FM Kitaev honeycomb model and only predicts a ferromagnetic ordered state, which is equivalent to a spin-wave theory. In the following, we would like to adopt a ferromagnetic order ansatz, which allows us to solve the mean-field theory using spin-wave theory in a more straightforward manner.
	
	\subsection{Spin-wave theory for the spin part Hamiltonian}
	
	Based on our DMRG results for $t>{}K$ and the findings of Schwinger boson mean-field theory, we would like to focus on the FM ordered phases. Instead of the Schwinger boson representation, for simplicity, here we use Holstein-Primakoff (HP) representation. For a semi-classical ferromagnetic order along the $(\sin\phi\cos\theta, \sin\phi\sin\theta, \cos\phi)$ direction, the ``spin operators'' (note that they are not the physical $S=1/2$ spins) now can be expressed as 
	\begin{flalign*}
		\hat{S}^x_j=&\Big{[}\cos\theta\sin\phi(\kappa-2a^\dag_ja^{}_j)+\cos\theta\cos\phi\sqrt{\kappa-a^\dag_j a^{}_j}(a^\dag_j+a^{}_j)\\
		&-\sin\theta\sqrt{\kappa-a^\dag_ja^{}_j}(ia^\dag_j-ia^{}_j)\Big{]}/2,&
	\end{flalign*}
	\begin{flalign*}
		\hat{S}^y_j=&\Big{[}\sin\theta\sin\phi(\kappa-2a^\dag_ja^{}_j)+\sin\theta\cos\phi\sqrt{\kappa-a^\dag_j a^{}_j}(a^\dag_j+a^{}_j)\\
		&+\cos\theta\sqrt{\kappa-a^\dag_ja^{}_j}(ia^\dag_j-ia^{}_j)\Big{]}/2,&
	\end{flalign*}
	\begin{flalign}
		\hat{S}^z_j=&\Big{[}\cos\phi(\kappa-2a^\dag_ja^{}_j)-\sin\phi\sqrt{\kappa-a^\dag_j a^{}_j}(a^\dag_j+a^{}_j)\Big{]}/2, &\label{eq:spinHP}
	\end{flalign}
	where $a^{\dag}_j$ and $a^{}_j$ are the creation and annihilation operators for the HP bosons (magnons), respectively. 
	
	\begin{figure}
		\includegraphics[width=0.5\linewidth]{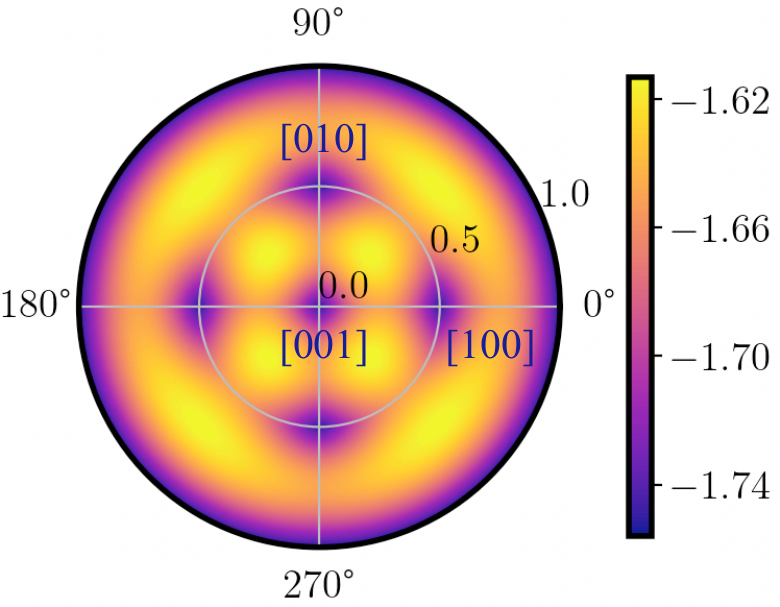}
		\caption{A colormap plot of the ground-state energy $E_s$ in Eq.~\eqref{eq:Es2} as a function of $\theta$ and $\phi$. The polar axis indicates $\theta\in[0,2\pi]$, and radius indicates $\phi\in[0,\pi]$. Here $\varGamma=0$.}\label{fig:gsG0}
	\end{figure}
	
	The Schwinger bosons are closely related to the HP bosons via their single occupancy constraints.  When $\theta=\phi=0$, it is easy to verify that by eliminating the $b^{}_{\uparrow}$ boson using the constraint,  the correspondence is~\cite{Auerbach1998} $$b^{}_{\downarrow}\leftrightarrow{}a^{}~~ {\rm  and }~~ b^{}_{\uparrow}\leftrightarrow{}\sqrt{\kappa-a^\dag a^{}}.$$ For the generic case, we can obtain that 
	\begin{equation}
		\begin{split}
			b_\uparrow{}&\rightarrow{}e^{-i\theta/2}\cos(\phi/2)\sqrt{\kappa-a^\dag a^{}}-e^{-i\theta/2}\sin(\phi/2)a^{},\\
			b_\downarrow{}&\rightarrow{}e^{i\theta/2}\sin(\phi/2)\sqrt{\kappa-a^\dag a^{}}+e^{i\theta/2}\cos(\phi/2)a^{}.
		\end{split}
	\end{equation} 
	We have $b^\dag_\uparrow{} b^{}_\uparrow+b^\dag_\downarrow{}b^{}_\downarrow=\kappa$ and hence $\kappa=n_b=1-\delta$ is the density of spinons. Naturally, the short-range ferromagnetic correlation, up to the second-order of HP bosons, becomes
	\begin{equation}
		\hat{F}^\gamma_{jk}=\sum_{s}\hat{F}^\gamma_{js,ks}=\kappa-\frac{a^\dag_ja^{}_j}{2}-\frac{a^\dag_ka^{}_k}{2}+a^\dag_{j}a^{}_k.~\label{eq:F2}
	\end{equation}

	Substituting Eq.~\eqref{eq:spinHP} into the spin part Hamiltonian, we can obtain a spin-wave mean-field theory up to the fourth-order:
	\begin{equation}
		\mathcal{H}_{s} = H^{(0)}_{s} + H^{(1)}_{s} + H^{(2)}_{s} + H^{(3)}_{s} + H^{(4)}_{s} + \mathcal{O}(b^5).
	\end{equation}
	Before discussing each term in the above Hamiltonian, it is worth emphasizing that the self-consistent conditions become Eqs.~\eqref{eq:P} and \eqref{eq:F2} which are numerically solved in an iterative manner.
	When carrying out the calculations, different randomly generated mean-field parameters are selected to initialize the iterations.

	Here $H^{(0)}_{s}$ is a constant term, namely, the semi-classical ground-state energy,
	\begin{equation*}
		\begin{split}
			H^{(0)}_{s}=&2\kappa\mathrm{N}_c\Re[{P_x+P_y+P_z}]t-N^2\mathrm{N}_cD[K+\varGamma{}f(\phi,\theta)]/4,
		\end{split}
	\end{equation*}
	where $\Re[\cdot]$ denotes taking the real part, $\mathrm{N}_c$ is the number of total unit cell, and $$f(\phi,\theta)=(\sin\theta+\cos\theta)\sin2\phi+\sin2\theta\sin^2\phi.$$
	Note that $D\equiv{}D_x=D_y=D_z$ is ensured by the $C_6$ lattice rotational symmetry.  
	The first-order term, $H^{(1)}_{s}$, is linear in the HP bosons:
	\begin{equation*}
		\begin{split}
			\frac{\kappa^{3/2}D\varGamma}{4}\sum_{j}\Big{\{}\Big{[}&(\sin\theta+\cos\theta)\cos2\phi+\sin2\theta\cos\phi\sin\phi+i(\cos\theta-\sin\theta)\cos\phi+i\cos2\theta\sin\phi\Big{]}a^\dag_{j}+h.c.\Big{\}}.
		\end{split}
	\end{equation*}
	Note that both $H^{(0)}_{s}$ and $H^{(1)}_{s}$ are independent of Kitaev coupling $K$, and thus, for $\varGamma=0$, there are no constrains on the values of $\phi$ and $\theta$. This indicates an emergent O(3) manifold of degenerate ground states for the pure Kitaev honeycomb model in the semiclassical limit. With $\varGamma>0$, a stable solution of ferromagnetic order requires $H^{(1)}_{s}$ to vanish, which, in turn, leads to a magnetic order along the [111] direction with $\phi=\arccos(\pm{}1/\sqrt{3})$ and $\theta=\pm\pi/4$.
	
	\begin{figure*}[!t]
		\includegraphics[width=\linewidth]{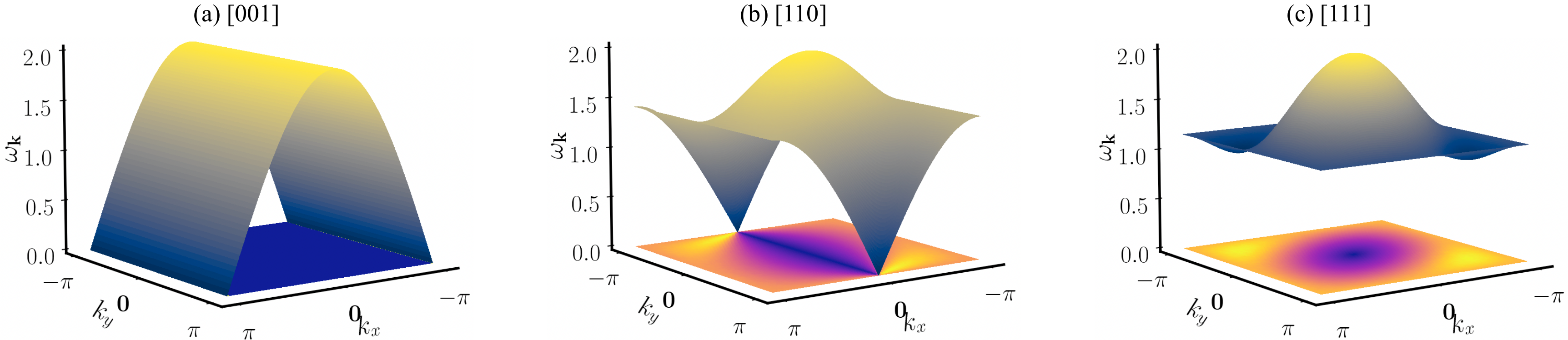}
		\caption{Magnon bands of $H^{(2)}_s$ in Eq.~\eqref{eq:Hs2} with $\varGamma=0$ for spin polarization in the (a) [001], (b) [110], and (c) [111] directions. Here we set the doping level $\delta=0$ and thereby $P_\gamma=0$. }\label{fig:msbare}
	\end{figure*}
	
	$H^{(2)}_s$ is the quadratic term in the Hamiltonian. In the basis of $\mathbb{B}^\dag_{{\bf k} }\equiv(a^\dag_{{\bf k},0}, a^\dag_{{\bf k}, 1}, a_{-{\bf k}, 0}, a_{-{\bf k}, 1})$, it reads
	\begin{equation}
		\begin{split}
			H^{(2)}_{s}=&\sum_{{\bf k}}\mathbb{B}^\dag_{{\bf k}}	
			\left( \begin{array}{cc}
				\mathbb{T}_{{\bf k}} & \mathbb{D}_{{\bf k}}  \\
				\mathbb{D}^\dag_{{\bf k}} & \mathbb{T}^T_{-{\bf k}}
			\end{array}\right)
			\mathbb{B}_{{\bf k}} + \mathrm{N}_c\Re[P_x+P_y+P_z]t - \frac{\kappa\mathrm{N}_cD{}}{2}[K+\varGamma{}f(\phi,\theta)],\label{eq:Hs2}
		\end{split}
	\end{equation}
	where $\mathbb{T}_{{\bf k}}$ and $\mathbb{D}_{{\bf k}}$ are $2\times{}2$ matrices:
	\begin{small}
		\begin{eqnarray}
			\mathbb{T}_{{\bf k}}=
			\frac{\kappa D}{4}\left(\begin{array}{r|r}
				& \frac{-K}{2}\left(\cos^2\theta\cos^2\phi+\sin^2\theta\right)+\frac{\varGamma}{2}\sin\theta\sin2\phi+ \\
				K+\varGamma{}f(\theta,\phi) & \left[\frac{-K}{2}\left(\sin^2\theta\cos^2\phi+\cos^2\theta\right)+ \frac{\varGamma}{2}\cos\theta\sin2\phi\right]e^{ik_x}+\\
				& \left[ \frac{K}{2}\sin^2\phi+\frac{\varGamma}{2}\sin2\theta\sin^2\phi \right]e^{ik_y}{~~~}\\
				\hline
				&  \\
				h.c.{\qquad\qquad}  &  K+\varGamma{}f(\theta,\phi)\\
				&  
			\end{array}\right)-
			\frac{t}{2}\left(\begin{array}{r|r}
				\Re[P_x]+	& -P_x^*+ \\
				\Re[P_y]+   & -P_y^*e^{-ik_x}+\\
				\Re[P_z]{~~~}     &   -P_z^*e^{-ik_y}{~~~}\\
				\hline
				& \Re[P_x]+ \\
				h.c.{\quad}  & \Re[P_y]+ \\
				& \Re[P_z]{~~~} 
			\end{array}\right),
		\end{eqnarray}
		and 
		\begin{equation}
			\mathbb{D}_{{\bf k}}=\frac{\kappa D}{4}\left(\begin{array}{r|r}
				& \frac{-K}{2}\left(\cos^2\theta\cos^2\phi-\sin^2\theta-i\sin2\theta\cos\phi \right)+ \\
				& \frac{-K}{2}\left(\sin^2\theta\cos^2\phi-\cos^2\theta+2i\sin2\theta\cos\phi\right)e^{ik_x}+ \\
				& -\frac{K}{2}\sin^2\phi{}e^{ik_y}+\\
				& \frac{\varGamma}{2}\left(\sin\theta\sin2\phi+2i\cos\theta\sin\phi\right)+\\
				& \frac{\varGamma}{2}\left(\cos\theta\sin2\phi-2i\sin\theta\sin\phi\right)e^{ik_x}+\\
				& -\frac{\varGamma}{2}\left[\sin2\theta(\cos^2\phi+1) + 2i\cos2\theta\cos\phi\right]e^{ik_y}{~~~}\\
				\hline
				\frac{-K}{2}\left(\cos^2\theta\cos^2\phi-\sin^2\theta-i\sin2\theta\cos\phi \right)+ &\\
				\frac{-K}{2}\left(\sin^2\theta\cos^2\phi-\cos^2\theta+2i\sin2\theta\cos\phi\right)e^{-ik_x}+  & \\
				-\frac{K}{2}\sin^2\phi{}e^{-ik_y}+& \\
				\frac{\varGamma}{2}\left(\sin\theta\sin2\phi+2i\cos\theta\sin\phi\right)+& \\
				\frac{\varGamma}{2}\left(\cos\theta\sin2\phi-2i\sin\theta\sin\phi\right)e^{-ik_x}+& \\
				-\frac{\varGamma}{2}\left[\sin2\theta(\cos^2\phi+1) + 2i\cos2\theta\cos\phi\right]e^{-ik_y}{~~~}&
			\end{array}\right).
		\end{equation}
	\end{small}
	
	We emphasize that when $\varGamma>0$, the spin polarization direction must be fixed as $\phi=\arccos(\pm{}1/\sqrt{3})$ and $\theta=\pm\pi/4$. It is worth noting that due to the lattice and time-reversal symmetries, in practice, we have $P_x=\Re[P_x]=P_y=P_z$. Therefore, we verify from the form of $\mathbb{T}_{{\bf k}}$ that the hole kinetic terms can play a role similar to the ferromagnetic Heisenberg interactions in the spin-wave theory with $J\equiv-\langle{}P_x\rangle{}t/\kappa.$
	
	\begin{figure}
		\includegraphics[width=1\linewidth]{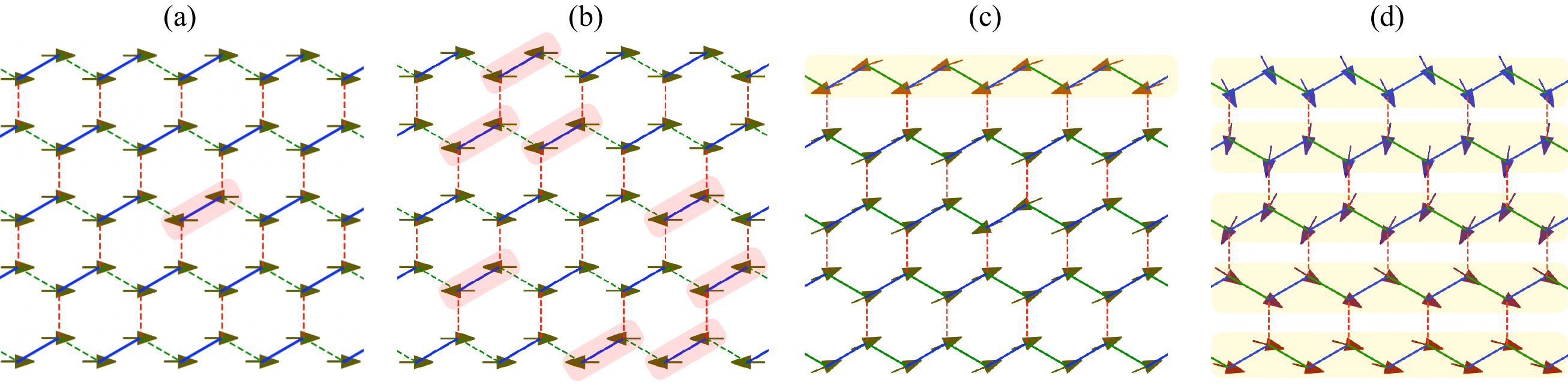}
		\caption{(a) In a [100] ordered ground state, where the effect of $z$-type (red dashed line) and $y$-type (green dashed line) bonds vanishes effectively, a zero-energy excitation can be constructed by flipping two spins (highlighted by the red shadow) connected by one $x$-type bonds (blue solid lines) into the [$\bar{1}00$] ($\bar{1}=-1$) direction. (b) Arbitrary numbers of zero-energy excitations can be created, leading to a flat magnon band for the [100] polarization. (c) In a [$xy$0] ordered ground state, the effect of $z$-type bonds (red dashed line) vanishes effectively and the honeycomb lattice is reduced into several independent chains formed by $x$- and $y$-type bonds (blue and green solid lines). The mechanism described in (a) and (b) no longer applies to this case. Instead, a zero-energy excitation can be created by rotating all the spins on a single chain (highlighted by the yellow shading) by an arbitrary angle around the $z$-axis. (d)  All the spins on each decoupled chain can be rotated freely around the $z$-axis without costing energies, leading to numerous zero-energy excitations and thereby a flat magnon bond for a [$xy$0] polarization. Note that the mechanism described by (c) and (d) is also applicable to the [100] ordered state.  }\label{fig:decoupledchains}
	\end{figure}

	\subsection{Order-by-disorder}
	Note that the constant term in Eq.~\eqref{eq:Hs2} arises from the zero-point energy contribution of the bosons, i.e., appears because we express this quadratic Hamiltonian in a bosonic BdG form.
	Therefore, the ground state energy of $\mathcal{H}_{s}$ up to the second order corrections reads
	\begin{equation}
		\begin{split}
			E_{s}=&\sum_{{\bf k}}\sum_{n=1,2}\omega_{{\bf k},n}+(2\kappa+1)\mathrm{N}_c\Re[P_x+P_y+P_z]t-\frac{\kappa(\kappa+2)\mathrm{N}_c}{4}[K+\varGamma{}f_0(\phi,\theta)],\label{eq:Es2}
		\end{split}
	\end{equation}
	where $\omega_{{\bf k},n}$ are the magnon dispersions. Note that $\omega_{{\bf k}, n}$ still depends on $\theta$ and $\phi$ even when $\varGamma=0$, indicating that the $O(3)$ degeneracy at the Kitaev point is lifted by quantum fluctuations. We find that a [001] (or equivalently [100] and [010]) magnetization is more energetically favorable than other directions, see Fig.~\ref{fig:gsG0}, indicating the presence of a quantum order-by-disorder mechanism.

	\begin{figure}
		\includegraphics[width=\linewidth]{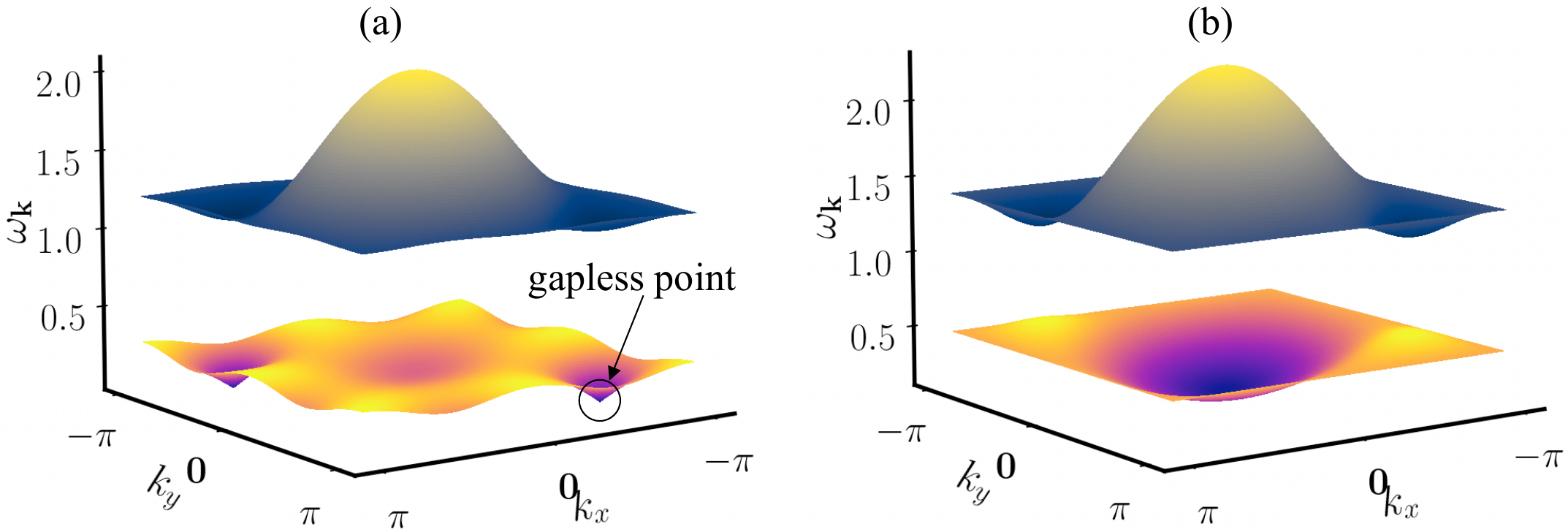}
		\caption{Magnon bands of $H^{(2)}_s$ in Eq.~\eqref{eq:Hs2} for spin polarization in the [111] direction. (a) $\varGamma=0.05$ and doping level $\delta=0$. (b)  $\varGamma=0.05$, $t=10$, and $\delta=0.01$, leading to a self-consistent solution of $P_{x,y,z}=-9.33\times{}10^{-3}$ and $F^{x,y,z}=0.86$. }\label{fig:msbare2}
	\end{figure}
	
	\subsection{Flat magnon band in $H^{(2)}_s$}
	
	When we diagonalize $H^{(2)}_s$, we find that the lower magnon band, $\omega_{{\bf k},0}$, is a nearly flat band with almost zero energy, see Figs.~\ref{fig:msbare}, regardless of the spin polarization directions. This flat band arises because the semiclassical ground state of the ferromagnetic Kitaev honeycomb model is not necessarily a ferromagnetic order but instead a classical spin liquid. In order words, one can construct numerous zero-energy excitations on top of a ferromagnetically ordered ground state in the semiclassical Kitaev honeycomb model, as demonstrated in Fig.~\ref{fig:decoupledchains}.  Though we use the [100] and [$xy$0] order as examples to illustrate the zero-energy excitations, the same mechanism is applicable for arbitrary [$xyz$] order. Indeed, one can construct the same type of excitations on top of a [$xyz$] order by fixing the $z$ components of all spins uniform and unchanged. Therefore, one can always observe flat magnon bands for arbitrary $[xyz]$ ordered ground state in the pure Kitaev honeycomb model.
	
	The flat band we discussed above can gain dispersion in two scenarios: (i) a finite positive $\varGamma$ which also forces the polarization direction to be [111], and (ii) a large $t$ with a finite doping level $\delta$ (i.e., a nonzero $P_\gamma$). Note that the lower band of $H^{(2)}_s$ exhibits gapless Goldstone-like modes even when $\varGamma>0$, as shown in Fig.~\ref{fig:msbare2}(a), although the system does not have any continuous symmetries to be spontaneously broken. This phenomenon has also been reported in Refs.~\cite{Joshi2018,Gohlke2020}. The zero-energy modes disappear in the presence of finite doping $\delta$, as shown in Fig.~\ref{fig:msbare2}(b).
	
	\subsection{Fourth-order correction to the magnon bands}
	The appearance of a flat band with zero energy and Goldstone-like modes indicates that $H^{(2)}_s$ is insufficient to capture the quantum fluctuations in the system. To properly account for these strong fluctuations, one must consider higher-order corrections from the Holstein-Primakoff expansion, specifically the quartic terms in the spin-wave Hamiltonian $H^{(4)}_s$ as (a positive $\varGamma>0$ will force the spin polarization to be fixed as $\phi=\arccos(\pm{}1/\sqrt{3})$ and $\theta=\pm\pi/4$)
	\begin{small}
		\begin{align*}
			H^{(4)}_{s}=\sum_{{\bf r}}&K\left[4\cos^2\theta\sin^2\phi{}a^\dag_{{\bf r},0}a^{}_{{\bf r},0}a^\dag_{{\bf r},1}a^{}_{{\bf r},1}+4\sin^2\theta\sin^2\phi{}a^\dag_{{\bf r}+x,0}a^{}_{{\bf r}+x,0}a^\dag_{{\bf r},1}a^{}_{{\bf r},1}+4\cos^2\phi{}a^\dag_{{\bf r}+y,0}a^{}_{{\bf r}+y,0}a^\dag_{{\bf r},1}a^{}_{{\bf r},1}\right]+\\
			&-\varGamma\left[4\sin\theta\sin2\phi{}a^\dag_{{\bf r},0}a^{}_{{\bf r},0}a^\dag_{{\bf r},1}a^{}_{{\bf r},1}+4\cos\theta\sin2\phi{}a^\dag_{{\bf r}+x,0}a^{}_{{\bf r}+x,0}a^\dag_{{\bf r},1}a^{}_{{\bf r},1}+4\sin2\theta\sin^2\phi{}a^\dag_{{\bf r}+y,0}a^{}_{{\bf r}+y,0}a^\dag_{{\bf r},1}a^{}_{{\bf r},1}\right]+\\
			&\left\{\left[\frac{K}{2}\left(\cos^2\theta\cos^2\phi+\sin^2\theta\right)-\frac{\varGamma}{2}\sin\theta\sin2\phi\right]\left(a^\dag_{{\bf r},0}a^{}_{{\bf r},0}a^\dag_{{\bf r},0}a^{}_{{\bf r},1}+a^\dag_{{\bf r},1}a^{}_{{\bf r},1}a^\dag_{{\bf r},1}a^{}_{{\bf r},0}\right)+h.c.\right\}+\\
			&\left\{\left[\frac{K}{2}\left(\cos^2\theta\cos^2\phi-\sin^2\theta-i\sin2\theta\cos\phi\right)-\frac{\varGamma}{2}\left(\sin\theta\sin2\phi+2i\cos\theta\sin\phi\right)\right]\left(a^\dag_{{\bf r},0}a^{}_{{\bf r},0}a^\dag_{{\bf r},0}a^\dag_{{\bf r},1}+a^\dag_{{\bf r},1}a^{}_{{\bf r},1}a^\dag_{{\bf r},1}a^\dag_{{\bf r},0}\right)+h.c.\right\}+\\
			&\left\{\left[\frac{K}{2}\left(\sin^2\theta\cos^2\phi+\cos^2\theta\right)-\frac{\varGamma}{2}\cos\theta\sin2\phi\right]\left(a^\dag_{{\bf r}+x,0}a^{}_{{\bf r}+x,0}a^\dag_{{\bf r}+x,0}a^{}_{{\bf r},1}+a^\dag_{{\bf r},1}a^{}_{{\bf r},1}a^\dag_{{\bf r},1}a^{}_{{\bf r}+x,0}\right)+h.c.\right\}+\\
			&\left\{\left[\frac{K}{2}\left(\sin^2\theta\cos^2\phi-\cos^2\theta+i\sin2\theta\cos\phi\right)-\frac{\varGamma}{2}\left(\cos\theta\sin2\phi-2i\sin\theta\sin\phi\right)\right]\left(a^\dag_{{\bf r}+x,0}a^{}_{{\bf r}+x,0}a^\dag_{{\bf r}+x,0}a^\dag_{{\bf r},1}+a^\dag_{{\bf r},1}a^{}_{{\bf r},1}a^\dag_{{\bf r},1}a^\dag_{{\bf r}+x,0}\right)+h.c.\right\}+\\
			&\left\{\left[\frac{K}{2}\sin^2\phi-\frac{\varGamma}{2}\sin2\theta\sin^2\phi\right]\left(a^\dag_{{\bf r}+y,0}a^{}_{{\bf r}+y,0}a^\dag_{{\bf r}+y,0}a^{}_{{\bf r},1}+a^\dag_{{\bf r},1}a^{}_{{\bf r},1}a^\dag_{{\bf r},1}a^{}_{{\bf r}+y,0}\right)+h.c.\right\}+\\
			&\left\{\left[\frac{K}{2}\sin^2\phi+\frac{\varGamma}{2} \left(\sin2\theta(\cos^2\phi+1) + 2i\cos2\theta\cos\phi \right)\right]
			\left(a^\dag_{{\bf r}+y,0}a^{}_{{\bf r}+y,0}a^\dag_{{\bf r}+y,0}a^\dag_{{\bf r},1}+a^\dag_{{\bf r},1}a^{}_{{\bf r},1}a^\dag_{{\bf r},1}a^\dag_{{\bf r}+y,0}\right)+h.c.\right\},
		\end{align*}
	\end{small}
	Note that in this context, we have neglected the fourth-order contributions from the kinetic terms since $P_\gamma\sim\delta$ is relatively small in the small doping limit. Furthermore, the cubic terms vanish for the [001] order. We also find that the correction of cubic terms are not important for the [111] order, also see Ref.~\cite{Chernyshev2009}.

	By introducing several Hartree-Fock parameters:
	\begin{equation*}
		\begin{array}{lll}
			n_l=\frac{1}{\mathrm{N}_c}\sum_{\bf k}\langle{}a^\dag_{{\bf k},l}a^{}_{{\bf k},l}\rangle,&
			\delta_l=\frac{1}{\mathrm{N}_c}\sum_{\bf k}\langle{}b_{{\bf k},l}a^{}_{{\bf k},l}\rangle, & \\
			m_x=\frac{1}{\mathrm{N}_c}\sum_{\bf k}\langle{}a^\dag_{{\bf k},0}a^{}_{{\bf k},1}\rangle,&
			m_y=\frac{1}{\mathrm{N}_c}\sum_{\bf k}e^{ik_x}\langle{}a^\dag_{{\bf k},0}a^{}_{{\bf k},1}\rangle, & m_z=\frac{1}{\mathrm{N}_c}\sum_{\bf k}e^{ik_y}\langle{}a^\dag_{{\bf k},0}a^{}_{{\bf k},1}\rangle,\\
			\Delta_x=\frac{1}{\mathrm{N}_c}\sum_{\bf k}\langle{}a^{}_{{\bf k},0}a^{}_{-{\bf k},1}\rangle, & 
			\Delta_y=\frac{1}{\mathrm{N}_c}\sum_{\bf k}e^{-ik_x}\langle{}a^{}_{{\bf k},0}a^{}_{-{\bf k},1}\rangle,&
			\Delta_z=\frac{1}{\mathrm{N}_c}\sum_{\bf k}e^{-ik_y}\langle{}a^{}_{{\bf k},0}a^{}_{-{\bf k},1}\rangle,\\
		\end{array}\label{eq:OPLWST}
	\end{equation*}
	the terms in $H^{(4)}_s$ on a $\alpha$-type bond $\langle ({\bf r},0)$, $({\bf r}',1) \rangle$ are decomposed (${\bf r}$ is the unit-cell index) as
	\begin{equation*}
		\begin{split}
			a^\dag_{{\bf r}, 0}a^{}_{{\bf r}, 0}a^\dag_{{\bf r}', 1}a^{}_{{\bf r}', 1}\rightarrow{}
			&n_{0}a^\dag_{{\bf r}', 1}a^{}_{{\bf r}', 1}+n_{1}a^\dag_{{\bf r}, 0}a^{}_{{\bf r}, 0}+n_{1}n_{1}+
			\Delta_\alpha{}a^\dag_{{\bf r}, 0}a^\dag_{{\bf r}',1}+\Delta_{\alpha}^*a^{}_{{\bf r}, 0}a^{}_{{\bf r}', 1}+|\Delta_\alpha|^2+ 
			m_\alpha{}a^\dag_{{\bf r}', 1}a^{}_{{\bf r}, 0}+m_{\alpha}^*a^\dag_{{\bf r}, 0}a^{}_{{\bf r}', 1}+|m_\alpha|^2,
			\\
			a^\dag_{{\bf r}, 0}a^{}_{{\bf r}, 0}a^\dag_{{\bf r}, 0}a^{}_{{\bf r}', 1}\rightarrow{}
			&2n_{0}a^\dag_{{\bf r}, 0}a^{}_{{\bf r}', 1}+m_\alpha a^\dag_{{\bf r}, 0}a^{}_{{\bf r}, 0}+
			m_\alpha a^{}_{{\bf r}, 0}a^\dag_{{\bf r}, 0}+2n_{0}m_\alpha+\delta_{0}^*a^{}_{{\bf r},0}a^{}_{{\bf r}',1}+\Delta_{\alpha}a^\dag_{{\bf r},0}a^\dag_{{\bf r},0}+\delta_{0}^*\Delta_\alpha,
			\\
			a^\dag_{{\bf r}', 1}a^{}_{{\bf r}', 1}a^\dag_{{\bf r}', 1}a^{}_{{\bf r}, 0}\rightarrow{}
			&2n_{1}a^\dag_{{\bf r}', 1}a^{}_{{\bf r}, 0}+m_{\alpha}^* a^\dag_{{\bf r}', 1}a^{}_{{\bf r}',1}+m_{\alpha}^* a^{}_{{\bf r}', 1}a^\dag_{{\bf r}', 1}+
			2n_{1}m_{\alpha}^*+
			\delta_{1}^*a^{}_{{\bf r}',1}a^{}_{{\bf r},0}+\Delta_{\alpha}a^\dag_{{\bf r}',1}a^\dag_{{\bf r}',1}+\delta_{1}^*\Delta_\alpha,
			\\
			a^\dag_{{\bf r}, 0}a^{}_{{\bf r}, 0}a^\dag_{{\bf r}, 0}a^\dag_{{\bf r}', 1}\rightarrow{}
			&2n_{0}a^\dag_{{\bf r}, 0}a^\dag_{{\bf r}', 1}+\Delta_{\alpha}^*a^\dag_{{\bf r}, 0}a^{}_{{\bf r}, 0}+\Delta_{\alpha}^*a^{}_{{\bf r}, 0}a^\dag_{{\bf r}, 0}+
			2n_{0}\Delta_{\alpha}^*+\delta_{0}^*a^\dag_{{\bf r}',1}a^{}_{{\bf r},0}+m_{\alpha}^*a^\dag_{{\bf r},0}a^\dag_{{\bf r},0}+\delta_{0}^*m_{\alpha}^*,\\  
			a^\dag_{{\bf r}', 1}a^{}_{{\bf r}', 1}a^\dag_{{\bf r}', 1}a^\dag_{{\bf r}, 0}\rightarrow{}
			&2n_{1}a^\dag_{{\bf r}', 1}a^\dag_{{\bf r}, 0}+\Delta_{\alpha}^*a^\dag_{{\bf r}', 1}a^{}_{{\bf r}', 1}+\Delta_{\alpha}^*a^{}_{{\bf r}', 1}a^\dag_{{\bf r}', 1}+
			2n_{1}\Delta_{\alpha}^*+\delta_{1}^*a^\dag_{{\bf r},0}a^{}_{{\bf r}',1}+m_{\alpha}a^\dag_{{\bf r}',1}a^\dag_{{\bf r}',1}+\delta_{1}^*m_\alpha.\\  
		\end{split}
	\end{equation*}
	Then the quartic interaction terms can be replaced with
	\begin{equation}
		H^{(4)}_s=\delta{}E^{(4)}_s+\delta{}H^{(2)}_{s}+...,
	\end{equation}
	where $\delta{}E^{(4)}_s$ and $\delta{}H^{(2)}_{s}$ are the Hartree-Fock corrections to the ground-state energy and magnon spectrum, respectively. 
	
	\begin{figure*}[!ht]
		\includegraphics[width=\linewidth]{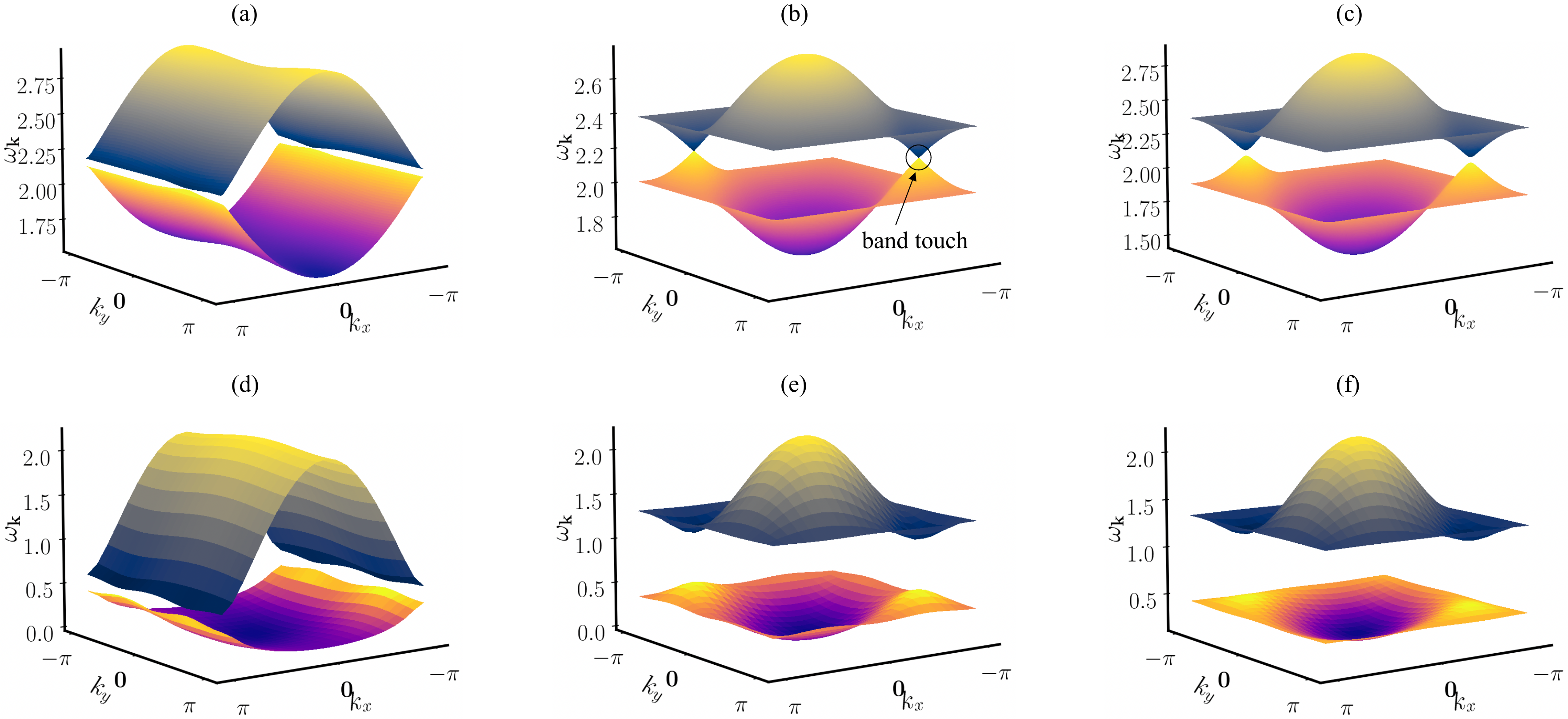}
		\caption{(a,b,c) Magnon bands of $H^{(2)}_s+\delta{}H^{(2)}_s$  [see corresponding definitions in Eqs.~\eqref{eq:Hs2} and \eqref{eq:Hs24}] (a) for spin polarization in the [001] direction with $\varGamma=0$, (b) for that in the [111] direction with $\varGamma=0$, and (c) for that in the [111] direction with $\varGamma=0.05$. Here we set $t=10$ and $\delta=0.01$, leading to self-consistent solutions of (a) $P_{x,y,z}=-9.17\times{}10^{-3}$, $F^x=F^y=0.79$, and $F^z=0.82$, (b) $P_{x,y,z}=-9.17\times{}10^{-3}$ and $F^{x,y,z}=0.83$, and (c) $P_{x,y,z}=-9.33\times{}10^{-3}$ and $F^{x,y,z}=0.86$. (d,e,f) Magnon 
			bands of $H^{(2)}$ only, for comparison. The model parameters and spin polarizations of figures in the same column are the same.}\label{fig:ms4}
	\end{figure*}
	
	Substituting the above decomposition into the quartic terms in $H^{(4)}_s$ leads to
	\begin{equation}
		\delta{}H^{(2)}_{s}=\sum_{{\bf k}}\mathbb{B}^\dag_{{\bf k}}\left( \begin{array}{cc}
			\delta{}\mathbb{T}_{{\bf k}} & \delta{}\mathbb{D}_{{\bf k}}  \\
			\delta{}\mathbb{D}^\dag_{{\bf k}} & \delta{}\mathbb{T}^T_{-{\bf k}}
		\end{array}\right)\mathbb{B}_{{\bf k}}.\label{eq:Hs24}
	\end{equation}
	Here the explicit form of $\delta\mathbb{T}_{\bf k}$ is 
	\begin{small}
		\begin{align}
			\delta\mathbb{T}_{{\bf k}}=&D\left(\begin{array}{r|r}
				2n_1K+	& 2m^*_x\left(K\cos^2\theta\sin^2\phi-\varGamma\sin\theta\sin2\phi\right)+ \\
				-2n_1\varGamma(\sin\theta+\cos\theta)\sin2\phi+   & 2m^*_y\left[K\left(\sin^2\theta\sin^2\phi\right) -\varGamma\cos\theta\sin2\phi\right]e^{ik_x}+\\
				-2n_1\varGamma\sin2\theta\sin^2\phi{~~~}  & 2m_z^*\left[ K\cos^2\phi-\varGamma\sin2\theta\sin^2\phi \right]e^{ik_y}{~~~}\\
				\hline
				& 2n_0K+ \\
				h.c.{\qquad\qquad}  & -2n_0\varGamma(\sin\theta+\cos\theta)\sin2\phi+ \\
				& -2n_0\varGamma\sin2\theta\sin^2\phi{~~~} 
			\end{array}\right)+\\
			\notag& D\left(\begin{array}{r|r}
				(m_x+m_x^*)\left[\frac{K}{2}\left(\cos^2\theta\cos^2\phi+\sin^2\theta\right)-\frac{\varGamma}{2}\sin\theta\sin2\phi\right]	& (n_0+n_1)\left[\frac{K}{2}\left(\cos^2\theta\cos^2\phi+\sin^2\theta\right)-\frac{\varGamma}{2}\sin\theta\sin2\phi\right]+ \\
				(m_y+m_y^*)\left[\frac{K}{2}\left(\sin^2\theta\cos^2\phi+\cos^2\theta\right)-\frac{\varGamma}{2}\cos\theta\sin2\phi\right]+   & (n_0+n_1)\left[\frac{K}{2}\left(\sin^2\theta\cos^2\phi+\cos^2\theta\right)-\frac{\varGamma}{2}\cos\theta\sin2\phi\right]e^{ik_x}+\\
				(m_z+m_z^*)\left[\frac{K}{2}\sin^2\phi-\frac{\varGamma}{2}\sin2\theta\sin^2\phi\right]{~~~}  & (n_0+n_1)\left[\frac{K}{2}\sin^2\phi-\frac{\varGamma}{2}\sin2\theta\sin^2\phi\right]e^{ik_y}{~~~}\\
				\hline
				& (m_x+m_x^*)\left[\frac{K}{2}\left(\cos^2\theta\cos^2\phi+\sin^2\theta\right)-\frac{\varGamma}{2}\sin\theta\sin2\phi\right]+ \\
				h.c.{\qquad\qquad}  & (m_y+m_y^*)\left[\frac{K}{2}\left(\sin^2\theta\cos^2\phi+\cos^2\theta\right)-\frac{\varGamma}{2}\cos\theta\sin2\phi\right]+ \\
				& (m_z+m_z^*)\left[\frac{K}{2}\sin^2\phi-\frac{\varGamma}{2}\sin2\theta\sin^2\phi\right]{~~~} 
			\end{array}\right)+\\
			\notag& D\left(\begin{array}{r|r}
				(\Delta_x^*+\Delta_x)\left[\frac{K}{2}\left(\cos^2\theta\cos^2\phi-\sin^2\theta\right)-\frac{\varGamma}{2}\sin\theta\sin2\phi\right]+	
				&
				\frac{\delta_1^*+\delta_0}{2}\left[\frac{K}{2}\left(\cos^2\theta\cos^2\phi-\sin^2\theta\right)-\frac{\varGamma}{2}\sin\theta\sin2\phi\right]+ \\
				(\Delta_x^*-\Delta_x)\left[-i\frac{K}{2}\sin2\theta\cos\phi-i{\varGamma}\cos\theta\sin\phi\right]+ 
				& \frac{\delta_1^*-\delta_0}{2}\left[-i\frac{K}{2}\sin2\theta\cos\phi-i{\varGamma}\cos\theta\sin\phi\right]+ \\
				(\Delta_y^*+\Delta_y)\left[\frac{K}{2}\left(\sin^2\theta\cos^2\phi-\cos^2\theta\right)-\frac{\varGamma}{2}\cos\theta\sin2\phi\right]+   & \frac{\delta_1^*+\delta_0}{2}\left[\frac{K}{2}\left(\sin^2\theta\cos^2\phi-\cos^2\theta\right)-\frac{\varGamma}{2}\cos\theta\sin2\phi\right]e^{ik_x}+\\
				(\Delta_y^*-\Delta_y)\left[i\frac{K}{2}\sin2\theta\cos\phi+i\varGamma\sin\theta\sin\phi\right]+ & \frac{\delta_1^*-\delta_0}{2}\left[i\frac{K}{2}\sin2\theta\cos\phi+i\varGamma\sin\theta\sin\phi\right]e^{ik_x}+
				\\ 
				(\Delta_z+\Delta_z^*)\left[\frac{K}{2}\sin^2\phi+\frac{\varGamma}{2}\sin2\theta(\cos^2\phi+1)\right]+ 
				& 
				\frac{\delta_0+\delta_1^*}{2}\frac{K}{2}\sin^2\phi{}e^{ik_y}+\\
				i(\Delta_z^*-\Delta_z)\varGamma\cos2\theta\cos\phi{~~~} & i\frac{\delta_1^*-\delta_0}{2}\varGamma\cos2\theta\cos\phi{}e^{ik_y}{~~~}
				\\ 
				\hline
				& (\Delta_x^*+\Delta_x)\left[\frac{K}{2}\left(\cos^2\theta\cos^2\phi-\sin^2\theta\right)-\frac{\varGamma}{2}\sin\theta\sin2\phi\right]+\\
				& 	(\Delta_x^*-\Delta_x)\left[-i\frac{K}{2}\sin2\theta\cos\phi-i{\varGamma}\cos\theta\sin\phi\right]+  \\
				h.c.{\qquad\qquad}  & (\Delta_y^*+\Delta_y)\left[\frac{K}{2}\left(\sin^2\theta\cos^2\phi-\cos^2\theta\right)-\frac{\varGamma}{2}\cos\theta\sin2\phi\right]+ \\
				& (\Delta_y^*-\Delta_y)\left[i\frac{K}{2}\sin2\theta\cos\phi+i\varGamma\sin\theta\sin\phi\right]+\\
				& (\Delta_z+\Delta_z^*)\left[\frac{K}{2}\sin^2\phi+\frac{\varGamma}{2}\sin2\theta(\cos^2\phi+1)\right]+\\
				& i(\Delta_z^*-\Delta_z)\varGamma\cos2\theta\cos\phi{~~~} 
			\end{array}\right).
		\end{align}
	\end{small}
	Similarly, 
	\begin{small}
		\begin{align}
			\delta\mathbb{D}_{{\bf k}}=&D\left(\begin{array}{r|r}
				& 2\Delta_x\left[ K\cos^2\theta\sin^2\phi -\varGamma\sin\theta\sin2\phi\right]+ \\
				& 2\Delta_y\left[K\sin^2\theta\sin^2\phi -\varGamma\cos\theta\sin2\phi \right]e^{ik_x}+ \\
				& 2\Delta_z\left[K\cos^2\phi{}-\varGamma\sin2\theta\sin^2\phi\right]e^{ik_y}+\\
				\hline
				2\Delta_x\left[ K\cos^2\theta\sin^2\phi -\varGamma\sin\theta\sin2\phi\right]+ &\\
				2\Delta_y\left[K\sin^2\theta\sin^2\phi -\varGamma\cos\theta\sin2\phi \right]e^{-ik_x}+  & \\
				2\Delta_z\left[K\cos^2\phi{}-\varGamma\sin2\theta\sin^2\phi\right]e^{-ik_y}+& \\
			\end{array}\right)+\\
			\notag&D\left(\begin{array}{r|r}
				(\Delta_x+\Delta_x^*)\left[\frac{K}{2}\left(\cos^2\theta\cos^2\phi+\sin^2\theta\right)-\frac{\varGamma}{2}\sin\theta\sin2\phi\right] & \frac{\delta_0+\delta_{1}}{2}\left[\frac{K}{2}\left(\cos^2\theta\cos^2\phi+\sin^2\theta\right)-\frac{\varGamma}{2}\sin\theta\sin2\phi\right]+ \\
				(\Delta_y+\Delta_y^*)\left[\frac{K}{2}\left(\sin^2\theta\cos^2\phi+\cos^2\theta\right)-\frac{\varGamma}{2}\cos\theta\sin2\phi\right]+ & \frac{\delta_0+\delta_{1}}{2}\left[\frac{K}{2}\left(\sin^2\theta\cos^2\phi+\cos^2\theta\right)-\frac{\varGamma}{2}\cos\theta\sin2\phi\right]e^{ik_x}+ \\
				(\Delta_z+\Delta_z^*)\left[\frac{K}{2}\sin^2\phi-\frac{\varGamma}{2}\sin2\theta\sin^2\phi\right]{~~~} & \frac{\delta_0+\delta_{1}}{2}\left[\frac{K}{2}\sin^2\phi-\frac{\varGamma}{2}\sin2\theta\sin^2\phi\right]e^{ik_y}{~~~}\\
				\hline
				\frac{\delta_0+\delta_{1}}{2}\left[\frac{K}{2}\left(\cos^2\theta\cos^2\phi+\sin^2\theta\right)-\frac{\varGamma}{2}\sin\theta\sin2\phi\right]+ & (\Delta_x+\Delta_x^*)\left[\frac{K}{2}\left(\cos^2\theta\cos^2\phi+\sin^2\theta\right)-\frac{\varGamma}{2}\sin\theta\sin2\phi\right]\\
				\frac{\delta_0+\delta_{1}}{2}\left[\frac{K}{2}\left(\sin^2\theta\cos^2\phi+\cos^2\theta\right)-\frac{\varGamma}{2}\cos\theta\sin2\phi\right]e^{-ik_x}+  & (\Delta_y+\Delta_y^*)\left[\frac{K}{2}\left(\sin^2\theta\cos^2\phi+\cos^2\theta\right)-\frac{\varGamma}{2}\cos\theta\sin2\phi\right]+ \\
				\frac{\delta_0+\delta_{1}}{2}\left[\frac{K}{2}\sin^2\phi-\frac{\varGamma}{2}\sin2\theta\sin^2\phi\right]e^{-ik_y}{~~~} &(\Delta_z+\Delta_z^*)\left[\frac{K}{2}\sin^2\phi-\frac{\varGamma}{2}\sin2\theta\sin^2\phi\right]{~~~} \\
			\end{array}\right)+\\
			&D\left(
			\begin{array}{r|r}
				m^*_x\frac{K}{2}\left(\cos^2\theta\cos^2\phi-\sin^2\theta-i\sin2\theta\cos\phi\right)+
				&
				(n_0+n_1)\frac{K}{2}\left(\cos^2\theta\cos^2\phi-\sin^2\theta-i\sin2\theta\cos\phi\right)+
				\\
				-m^*_x\frac{\varGamma}{2}\left(\sin\theta\sin2\phi+2i\cos\theta\sin\phi\right)+
				&
				-(n_0+n_1)\frac{\varGamma}{2}\left(\sin\theta\sin2\phi+2i\cos\theta\sin\phi\right)+
				\\
				m^*_y\frac{K}{2}\left(\sin^2\theta\cos^2\phi-\cos^2\theta+i\sin2\theta\cos\phi\right)+
				&
				(n_0+n_1)\frac{K}{2}\left(\sin^2\theta\cos^2\phi-\cos^2\theta+i\sin2\theta\cos\phi\right)e^{ik_x}+\\
				-m^*_y\frac{\varGamma}{2}\left(\cos\theta\sin2\phi-2i\sin\theta\sin\phi\right)+
				&
				-(n_0+n_1)\frac{\varGamma}{2}\left(\cos\theta\sin2\phi-2i\sin\theta\sin\phi\right)e^{ik_x}+
				\\	
				m^*_z\frac{K}{2}\sin^2\phi+
				&
				(n_0+n_1)\frac{K}{2}\sin^2\phi{}e^{ik_y}+\\
				m^*_z\frac{\varGamma}{2} \left(\sin2\theta(\cos^2\phi+1) + 2i\cos2\theta\cos\phi \right)  {~~~}
				&
				(n_0+n_1)\frac{\varGamma}{2} \left(\sin2\theta(\cos^2\phi+1) + 2i\cos2\theta\cos\phi \right)e^{ik_y}{~~~}\\
				\hline	
				(n_0+n_1)\frac{K}{2}\left(\cos^2\theta\cos^2\phi-\sin^2\theta-i\sin2\theta\cos\phi\right)+
				&
				m_x\frac{K}{2}\left(\cos^2\theta\cos^2\phi-\sin^2\theta-i\sin2\theta\cos\phi\right)+
				\\
				-(n_0+n_1)\frac{\varGamma}{2}\left(\sin\theta\sin2\phi+2i\cos\theta\sin\phi\right)+
				&
				-m_x\frac{\varGamma}{2}\left(\sin\theta\sin2\phi+2i\cos\theta\sin\phi\right)+
				\\
				(n_0+n_1)\frac{K}{2}\left(\sin^2\theta\cos^2\phi-\cos^2\theta+i\sin2\theta\cos\phi\right)e^{-ik_x}+
				&
				m_y\frac{K}{2}\left(\sin^2\theta\cos^2\phi-\cos^2\theta+i\sin2\theta\cos\phi\right)+
				\\
				-(n_0+n_1)\frac{\varGamma}{2}\left(\cos\theta\sin2\phi-2i\sin\theta\sin\phi\right)e^{-ik_x}+
				&
				-m_y\frac{\varGamma}{2}\left(\cos\theta\sin2\phi-2i\sin\theta\sin\phi\right)+
				\\	
				(n_0+n_1)\frac{K}{2}\sin^2\phi{}e^{-ik_y}+
				&
				m_z\frac{K}{2}\sin^2\phi+
				\\
				(n_0+n_1)\frac{\varGamma}{2} \left(\sin2\theta(\cos^2\phi+1) + 2i\cos2\theta\cos\phi \right)e^{-ik_y}{~~~}
				&
				m_z\frac{\varGamma}{2} \left(\sin2\theta(\cos^2\phi+1) + 2i\cos2\theta\cos\phi \right)  {~~~}
			\end{array}
			\right).
		\end{align}
	\end{small}
	And the fourth-order correction to the ground-state energy is 
	\begin{small}
		\begin{align}
			\delta{}E^{(4)}_s=\sum_{\mathbf{k}}&4\left[K-\varGamma(\sin\theta\sin2\phi+\cos\theta\sin2\phi+4\sin2\theta\sin^2\phi)\right]4n_0n_1+4\left(K\cos^2\theta\sin^2\phi{}-\varGamma\sin\theta\sin2\phi\right)\left(|m_x|^2+|\Delta_x|^2\right)+\notag\\
			&4\left(K\sin^2\theta\sin^2\phi{}-\varGamma\cos\theta\sin2\phi\right)\left(|m_y|^2+|\Delta_y|^2\right)+4\left(K\cos^2\phi{}-\varGamma\sin2\theta\sin^2\phi\right)\left(|m_z|^2+|\Delta_z|^2\right)+\notag\\
			&\left[\frac{K}{2}\left(\cos^2\theta\cos^2\phi+\sin^2\theta\right)-\frac{\varGamma}{2}\sin\theta\sin2\phi\right]\left(2n_0m_x+2n_1m_x^*+\delta_0^*\Delta_x+\delta_1^*\Delta_x+c.c.\right)+\notag\\
			&\left[\frac{K}{2}\left(\cos^2\theta\cos^2\phi+\sin^2\theta\right)-\frac{\varGamma}{2}\sin\theta\sin2\phi\right]\left(2n_0m_x+2n_1m_x^*+\delta_0^*\Delta_x+\delta_1^*\Delta_x+c.c.\right)+\notag\\
			&\left[\frac{K}{2}\sin^2\phi-\frac{\varGamma}{2}\sin2\theta\sin^2\phi\right]\left(2n_0m_z+2n_1m_z^*+\delta_0^*\Delta_z+\delta_1^*\Delta_z+c.c.\right)+\notag\\	
			&\left[\frac{K}{2}\sin^2\phi-\frac{\varGamma}{2}\sin2\theta\sin^2\phi\right]\left(2n_0m_z+2n_1m_z^*+\delta_0^*\Delta_z+\delta_1^*\Delta_z+c.c.\right)+\notag\\	
			&\left\{\left[\frac{K}{2}\left(\cos^2\theta\cos^2\phi-\sin^2\theta-i\sin2\theta\cos\phi\right)-\frac{\varGamma}{2}\left(\sin\theta\sin2\phi+2i\cos\theta\sin\phi\right)\right]\left(2n_0\Delta_x^*+2n_1\Delta_x^*+\delta_0^*m_x^*+\delta_1^*m_x\right)+c.c.\right\}+\notag\\
			&\left\{\left[\frac{K}{2}\left(\sin^2\theta\cos^2\phi-\cos^2\theta+i\sin2\theta\cos\phi\right)-\frac{\varGamma}{2}\left(\cos\theta\sin2\phi-2i\sin\theta\sin\phi\right)\right]\left(2n_0\Delta_y^*+2n_1\Delta_y^*+\delta_0^*m_y^*+\delta_1^*m_y\right)+c.c.\right\}+\notag\\
			&\left\{\left[\frac{K}{2}\sin^2\phi+\frac{\varGamma}{2} \left(\sin2\theta(\cos^2\phi+1) + 2i\cos2\theta\cos\phi\right)\right]\left(2n_0\Delta_z^*+2n_1\Delta_z^*+\delta_0^*m_z^*+\delta_1^*m_z\right) +c.c.\right\}\notag
		\end{align}
	\end{small}
	
	In the end, we would like to comment that the inclusion of $\delta H^{(2)}_s$ leads to several important effects in the spin-wave theory.
	In Fig~\ref{fig:ms4}, we show the magnon bands corrected by the fourth-order term $\delta H^{(2)}_s$. We find that $\delta H^{(2)}_s$ significantly changes the dispersions of both the lower and upper magnon bands. In particular, the lower band gains much more dispersion, as shown in Fig.~\ref{fig:ms4}(a) and (b). Furthermore, the energy gap between the lower and upper magnon bands becomes closed due to the fourth-order corrections  when $\varGamma$ vanishes, see Fig.~\ref{fig:ms4}(b). It indicates that the topological properties of the magnon bands are also changed by $\delta{}H^{(2)}_s$. Finally, a finite $\varGamma$ can reopen the energy gap, as shown in Fig.~\ref{fig:ms4}(c).
	
	\section*{Supplementary Note 4: Zigzag order} \label{app:zigzag}
	
	\begin{figure}
		\includegraphics[width=0.5\linewidth]{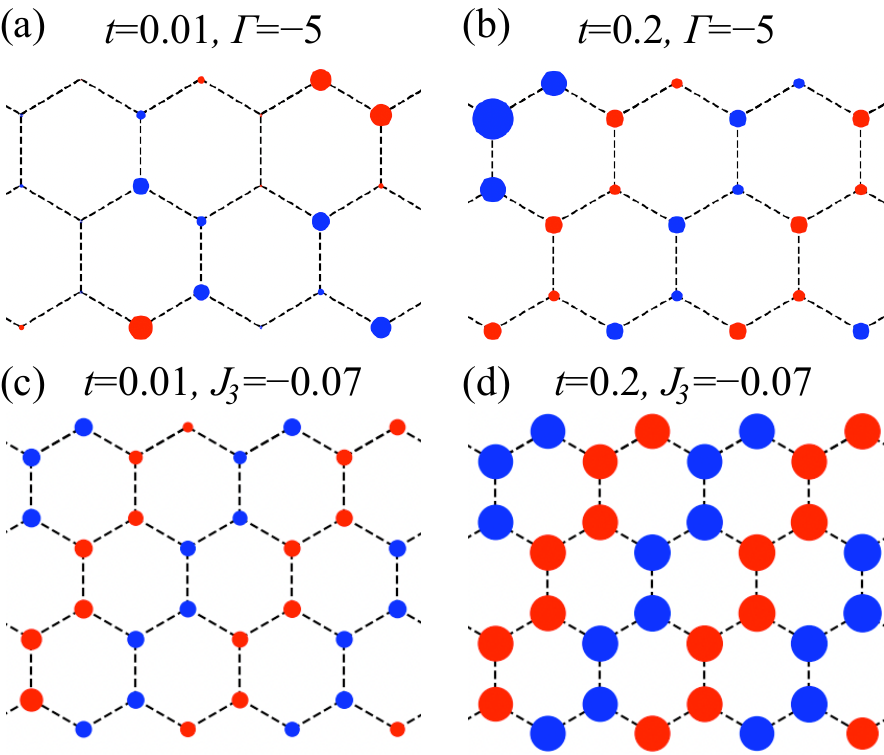}
		\caption{Zigzag order induced by mobile holes. Real-space magnetization $\langle{}S^z_j\rangle$ ($\langle{}S^x_j\rangle$ and $\langle{}S^y_j\rangle$ are similar) profiles for (a, b) $t$-$K$-$\varGamma$ model on YC3 cylinders and (c,d) $t$-$K$-$J_3$ model ($\varGamma=0$) on YC4 cylinders. Here $J_3$ is the 3rd NN antiferromagnetic Heisenberg exchange, $K=1$, and the doping level is $\delta\approx0.015$. Positive and negative magnetic moments are shown by red and blue dots, respectively, whose radius scales with the magnitude of magnetization $|\langle{}S^z_j\rangle|$, such that $|\langle{}S^z_j\rangle|\approx0.06$ in the two middle columns in (b).}\label{fig:zigzagmain}
	\end{figure}

	Motivated by the experimental Kitaev materials, we discuss the effect of hole doping on the zigzag ordered state. Previous studies~\cite{Rau2014,Janssen2017,Gohlke2018} report that the $K$-$\varGamma$ model alone is insufficient to support a zigzag order, even with a dominant antiferromagnetic $\varGamma<0$. Our simulations are consistent with these findings, as we observe something like spiral and/or incommensurate orders rather than zigzag order up to $\varGamma=-5K$, see Fig.~\ref{fig:zigzagmain}(a). On the other hand, the $t$-$K$-$\varGamma$ model at a doping level $\delta\approx0.01$ can stabilize the zigzag ordered phase with a moderate $t\approx{}0.2K$ with $\varGamma=-5K$, as shown in Fig.~\ref{fig:zigzagmain}(b). This is attributed to that the hole kinetic terms act as a ferromagnetic Heisenberg exchange effectively, $J\sim{}|t|\delta{}$, which can prompt a zigzag order in the $K$-$\varGamma$ model~\cite{Laubach2017}.
	
	We find that a relatively large value of $|\varGamma|>K$ is required to achieve a zigzag order in the $t$-$K$-$\varGamma$ model, which might be due to the usage of narrow cylindrical geometries in the DMRG simulations~\cite{Wang2019}. We anticipate that this value will significantly decrease in the two-dimensional limit. Alternatively, we introduce a small third NN antiferromagnetic Heisenberg exchange $J_3<0$ on top of the $t$-$K$ ($\varGamma=0$) model, which has shown to be sufficient to drive the system to a zigzag ordered phase~\cite{Winter2016,Winter2017NC,Catuneanu2018}, even on narrow cylinders. We find that in this scenario, a small $t$ can also enhance the stability of the zigzag phase. For instance, when $J_3=-0.07K$, a doping level of $\delta\approx{}0.01$ and a hopping strength of $t=0.2K$ can approximately double the magnitude of zigzag order compared to the case at $t\rightarrow{}0$. Overall, our results suggest that a zigzag order can be more easily stabilized in the $K$-$\varGamma$ ($K$-$J_3$) model by supplementing it with a moderate hole kinetic term $t$ as well as small hole doping. Nevertheless, it is worth noting that if the hopping strength $t$ dominates over the antiferromagnetic interactions of $\varGamma<0$ ($J_3<0$),  the system still undergoes a ferromagnetic ordering transition. 
	
	\section*{Supplementary Note 5: Spin-1/2 moment around $t=0$} \label{app:freeSpin}
	
	Here we clarify the absence of a free spin moment in the gapless phase even in the limit of a static hole, i.e.  hopping $t=0$. As discussed in Ref. [49] of the main text, the KHM with a vacancy is still exactly solvable with the help of the Majorana representation. Denoting three Pauli matrices as $\sigma^\gamma$ ($\gamma=x,y,z$), we introduce $\sigma^\gamma_j=ic^0_ic^\gamma_i$ with $c^0_i$ and $c^\gamma_i$ being the matter and gauge Majorana fermions, respectively. The doped hole introduces a static vacancy to the KHM. Without loss of generality, we assume that this vacancy is located at the site $i_0$ of $A$ sublattice. Then, there are three unpaired gauge Majoranas (namely, Majorana zero modes), $c_j^x$, $c_k^y$, and $c_l^z$, where $j$, $k$, and $l$ are three NN sites of $i_0$. 
	And there must be one Majorana zero mode $\gamma_0$ as a superposition of matter Majorana fermions. 
	
	\begin{figure}
		\includegraphics[width=0.5\linewidth]{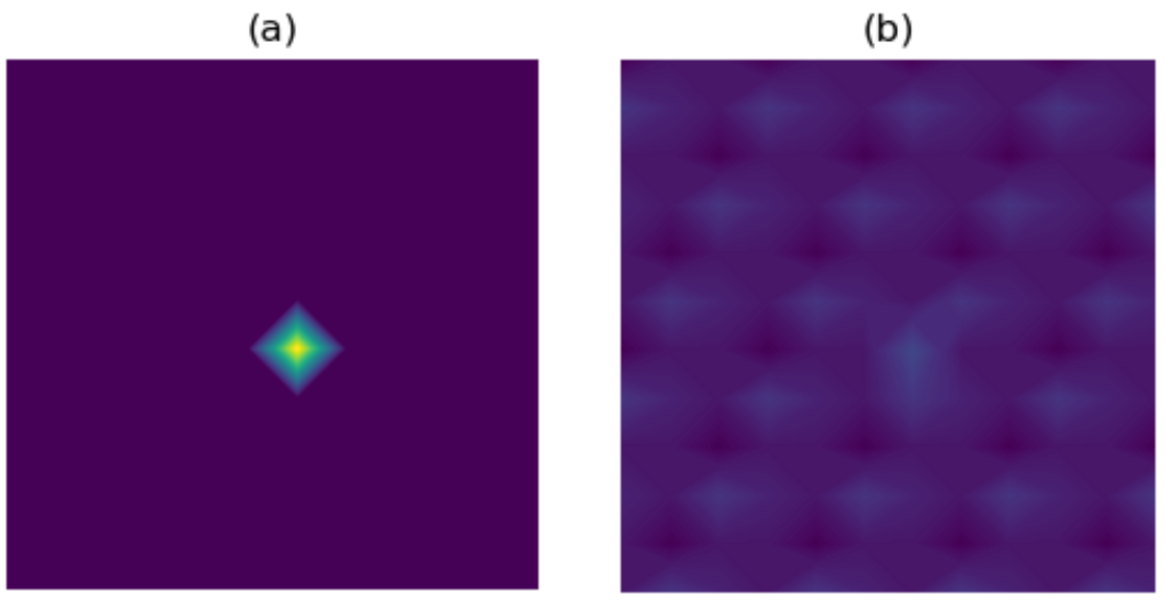}
		\caption{The single-particle wave function of the Majorana zero mode $\gamma_0$ in the matter sector on a $L_y=6$ and $L_x=12$ honeycomb lattice. Note that we plot the honeycomb lattice as a $2L_y\times L_x$ square lattice, where factor 2 accounts for the $A$ and $B$ sublattices. (a) The gapped A phase with $K_z=3K_x=3K_y$. (b) The gapless B phase with $K_z=K_x=K_y$. }\label{fig:S12}
	\end{figure}
	
	After a Gutzwiller projection those four Majorana zero modes can lead to two-fold ground-state degeneracy. Moreover, in the gapped A phase~\cite{Kitaev06}, $\gamma_0$ is spatially localized around the static vacancy, as shown in Fig.~\ref{fig:S12} (a). It indicates that in the doped A phase, those four modes can effectively form a {\it free and local} spin-1/2 moment, leading to a finite magnetization of a single free spin-1/2 in the ground state manifold. However, in the gapless B phase, $\gamma_0$ is no longer a local zero mode, see Fig.~\ref{fig:S12} (b). Therefore, although the system exhibits an odd number of spins and the ground states must be two-fold degenerate, there is not necessarily a {\it free} spin-1/2 moment. In stark contrast, this effective spin-$1/2$ degree of freedom, which exhibits zero-energy excitation, strongly interacts with the rest of the system. In this case, even though the ground state has an odd number of spins, it does not give rise to a magnetization of a free spin-1/2 moment.

\end{widetext}

\end{document}